\documentclass[11pt, a4paper]{article}

\usepackage[utf8]{inputenc}
\usepackage[english]{babel} 

\usepackage[top=2.5cm,bottom=2cm,right=2.5cm,left=3.5cm]{geometry}
\usepackage[onehalfspacing]{setspace}
\usepackage{fancyhdr}
\usepackage[frozencache]{minted}
\usepackage{listings}
\lstdefinestyle{xtext}{
columns=fixed,
breaklines=true,
numbers=left,
numberstyle=\tiny,
stepnumber=1,
numbersep=5pt,
showspaces=false,
showstringspaces=false,
frame=tblr,
tabsize=2,
captionpos=b,
frame=shadowbox,
columns=fixed,
basicstyle=\ttfamily\scriptsize,
rulesepcolor=\color[rgb]{0.6, 0.6, 0.6},
keywordstyle=\color[rgb]{0.5,0,0.34}\textbf,
stringstyle=\color{blue},
commentstyle=\color[rgb]{0.4,0.7,0.4},
backgroundcolor=\color[rgb]{0.97,0.97,0.97},
alsodigit={:},
morekeywords={
name,
url,
requestMethod,
query,
headers,
body,
returnValue,
customization,
type,
messages,
terminal,
fragment,
http,
key,
value,
enum
},
morestring=[b]",
morestring=[b]',
showtabs=false,
morecomment=[l]{//},
morecomment=[s]{/*}{*/}
}

\lstdefinestyle{HTTP}{
columns=fixed,
breaklines=true,
numbers=left,
numberstyle=\tiny,
stepnumber=1,
numbersep=5pt,
showspaces=false,
showstringspaces=false,
frame=tblr,
tabsize=2,
captionpos=b,
frame=shadowbox,
columns=fixed,
basicstyle=\ttfamily\scriptsize,
rulesepcolor=\color[rgb]{0.6, 0.6, 0.6},
keywordstyle=\color[rgb]{0.5,0,0.34}\textbf,
stringstyle=\color{blue},
backgroundcolor=\color[rgb]{0.97,0.97,0.97},
alsodigit={:},
morekeywords={
http,
name,
url,
server,
path,
type,
GET,
param,
input,
header,
body,
contentType,
payload,
entityType,
customize,
ProxyServer,
basicAuth,
timeout,
port,
username,
password
},
morestring=[b]",
showtabs=false
}

\lstdefinestyle{XtendTemplate}{
columns=fixed,
breaklines=true,
numbers=left,
numberstyle=\tiny,
stepnumber=1,
numbersep=5pt,
showspaces=false,
showstringspaces=false,
frame=tblr,
tabsize=2,
captionpos=b,
frame=shadowbox,
columns=fixed,
basicstyle=\ttfamily\scriptsize,
rulesepcolor=\color[rgb]{0.6, 0.6, 0.6},
keywordstyle=\color[rgb]{0.5,0,0.34}\textbf,
stringstyle=\color{blue},
backgroundcolor=\color[rgb]{0.97,0.97,0.97},
alsodigit={:},
morekeywords={
classname,
message
}
morestring=[b]",
showtabs=false
}

\lstdefinestyle{java}{
columns=fixed,
breaklines=true,
numbers=left,
numberstyle=\tiny,
stepnumber=1,
numbersep=5pt,
showspaces=false,
showstringspaces=false,
frame=tblr,
tabsize=2,
captionpos=b,
frame=shadowbox,
columns=fixed,
basicstyle=\ttfamily\scriptsize,
rulesepcolor=\color[rgb]{0.6, 0.6, 0.6},
keywordstyle=\color[rgb]{0.5,0,0.34}\textbf,
stringstyle=\color{blue},
backgroundcolor=\color[rgb]{0.97,0.97,0.97},
alsodigit={:},
morekeywords={
package,
import,
public,
class,
try,
throws,
new,
if,
finally,
return
}
morestring=[b]",
morestring=[b]',
showtabs=false
}

\lstdefinestyle{XtendTemplate}{
columns=fixed,
breaklines=true,
numbers=left,
numberstyle=\tiny,
stepnumber=1,
numbersep=5pt,
showspaces=false,
showstringspaces=false,
frame=tblr,
tabsize=2,
captionpos=b,
frame=shadowbox,
columns=fixed,
basicstyle=\ttfamily\scriptsize,
rulesepcolor=\color[rgb]{0.6, 0.6, 0.6},
keywordstyle=\color[rgb]{0.5,0,0.34}\textbf,
stringstyle=\color{blue},
backgroundcolor=\color[rgb]{0.97,0.97,0.97},
alsodigit={:},
morekeywords={
classname,
message
}
morestring=[b]",
showtabs=false
}

\lstdefinestyle{java}{
columns=fixed,
breaklines=true,
numbers=left,
numberstyle=\tiny,
stepnumber=1,
numbersep=5pt,
showspaces=false,
showstringspaces=false,
frame=tblr,
tabsize=2,
captionpos=b,
frame=shadowbox,
columns=fixed,
basicstyle=\ttfamily\scriptsize,
rulesepcolor=\color[rgb]{0.6, 0.6, 0.6},
keywordstyle=\color[rgb]{0.5,0,0.34}\textbf,
stringstyle=\color{blue},
backgroundcolor=\color[rgb]{0.97,0.97,0.97},
alsodigit={:},
morekeywords={
package,
import,
public,
class,
try,
throws,
new,
if,
final,
return,
switch,
case,
default,
package,
@Override,
else,
catch,
break,
throw,
private,
finally
}
morestring=[b]",
morestring=[b]',
showtabs=false
}

\usepackage{pdfpages}

\usepackage{hyperref}

\addto\extrasenglish{

}

\usepackage{todonotes}

\usepackage{graphicx}
\usepackage{enumitem}
\usepackage{tabularx}
\usepackage{booktabs}
\usepackage[all]{hypcap} 
\usepackage{tocbibind}
\usepackage[toc, nonumberlist]{glossaries}
\usepackage[toc, titletoc, title, page]{appendix}
\usepackage{tikz} 
\def\checkmark{\tikz\fill[scale=0.4](0,.35) -- (.25,0) -- (1,.7) -- (.25,.15) -- cycle;} 

\usepackage[backend=biber]{biblatex}
\usepackage{csquotes}

\newcommand{\frontmatter}{
	\cleardoublepage
	\pagenumbering{Roman}
}

\newcommand{\classname}[1]{\texttt{#1}}

\newcommand{\mainmatter}{
	\cleardoublepage
	\pagenumbering{arabic}
	\setcounter{page}{1}
}
\newcommand{\attachmentmatter}{
	\cleardoublepage
	\pagenumbering{Roman}
	\setcounter{page}{1}
}

\begin{document}

\newgeometry{top=0.7cm,bottom=2cm,right=2cm,left=2cm}
\graphicspath{{./title_page/title_page_figures/}}

\begin{titlepage}
{
\par
\raisebox{\height}{\includegraphics[width=3.3cm]{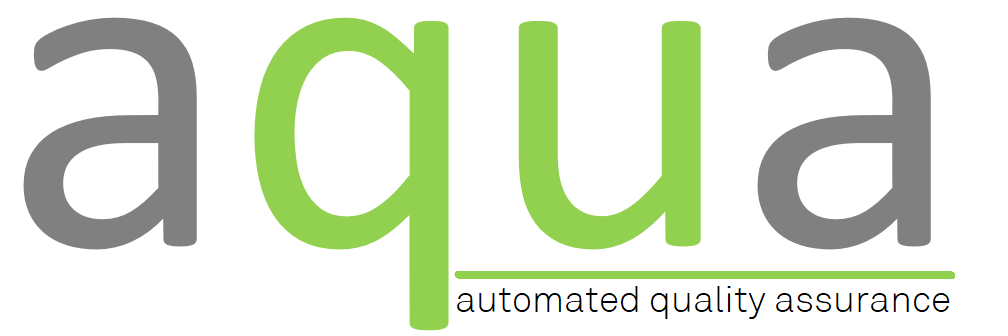}}
\hfill
\raisebox{\height}{\includegraphics[width=3.8cm]{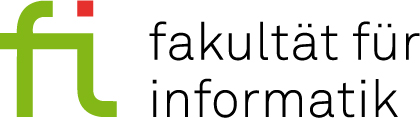}}
\hfill
\raisebox{2.25\height}{\includegraphics[width=4.1cm]{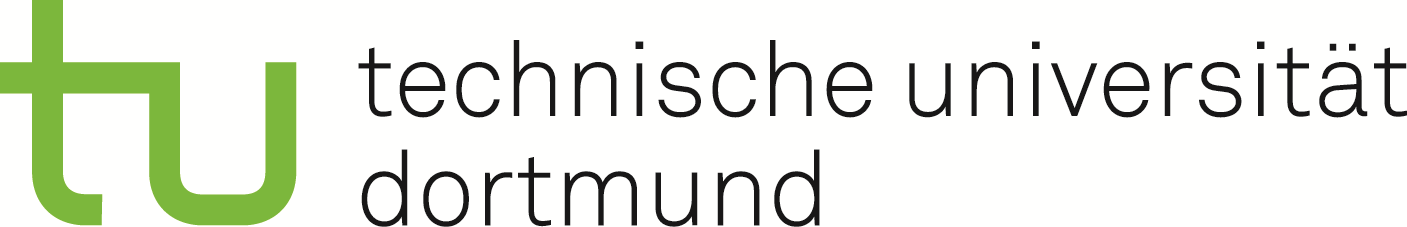}}
\par
\vspace{-30pt}

\small
\begin{tabularx}{\textwidth}{ XXX }
	\toprule
	\bigbreak \\
	Lehrstuhl für & Fakultät Informatik & Technische Universität Dortmund \\
	Software Engineering & & \\
	Prof. Dr. Falk Howar & & \\
\end{tabularx}
\vspace{4cm}

\begin{center}
\Huge
Bachelorarbeit \\
\vspace{3cm}
\Large

DSL-driven Integration of HTTP Services in DIME\\

\vspace{6cm}
\large
bearbeitet von: Bruno Steffen
\bigbreak
\normalsize
Studiengang: Informatik\\
\bigbreak
Erstgutachter: Prof. Dr. Falk Howar\\
Zweitgutachter: Dr. Stefan Naujokat\\
Betreuer: Jonas Sch\"urmann
\end{center}

}
\end{titlepage}
\restoregeometry
\newpage

\frontmatter
\tableofcontents
\newpage

\mainmatter
\section{Introduction}
\label{sec:intro}

Web applications are applications made available to users globally via the World Wide Web (WWW). They are growing in popularity compared to traditional client-server applications with limited access. Modern day web applications also have increasingly large user bases, which in turn require scalable software architectures.\\
Web services are an extension to the web application model. The W3C~\cite{w3} writes in their definition of a web service that 
\begin{quote}
"A Web service is a software system designed to support interoperable machine-to-machine interaction over a network." \cite{webService}
\end{quote} 
This means that, in contrast to web applications, web services are not designed for global access by users, but by applications. Hence, the demand for web services has been growing side by side with the steadily increasing amount of web applications worldwide. 
This trend has two effects:
\begin{enumerate}
    \item Developers have to offer their services in shape of web services to make them available globally.
    \item More and more web services exist that could be accessed globally for various web applications to use.
\end{enumerate}
Consequently, it became increasingly important for web developers to be able to access and supply web services.\\
DIME~\cite{BFKLNN2016} is a tool that promises a model driven approach to creating web applications, thus making it easier to launch web applications for non-programmers. However, when implementing a new web application one often needs to include preexisting or custom web services to support certain features.

\paragraph{Accessing Web Services.}
Web services are often accessed through the Hypertext Transfer Protocol \ref{subsec:HTTP}, short HTTP.
HTTP is a protocol in the application layer of the OSI and TCP/IP models that describe communication systems and their functions. Almost 47\% \cite{HTTP1} of websites use the prevalent version HTTP/2 and another 23.8\% \cite{HTTP1} use the newest version HTTP/3, making it by far the most widely used protocol for websites and web services. Furthermore, HTTP is commonly used for the communication between microservices and has no compatibility issues with a wide range firewalls. Accordingly, it can be assumed that the most effective method for guaranteeing access to as many internal and external web services as possible is to use HTTP for communication.\\
For that reason, most of the applications that have already been made in DIME make use of HTTP, and upcoming projects will need to integrate it as well. \\

However, there is no native HTTP-Client functionality within DIME. Thus, every single application has a different implementation, using either the native Java HTTP Client or libraries that can further facilitate the developers' needs. The problem with this approach is that the same functionality is redesigned and reimplemented over and over again, creating a potential for multiple hazards. These could be discrepancies between different DIME web applications, duplications, higher maintenance need, and therefore clutter and waste in the sense of Poppendieck \cite{PoppendieckPoppendieck03}.\\
\begin{figure}[h]
	\begin{center}
		\includegraphics[width=\textwidth]{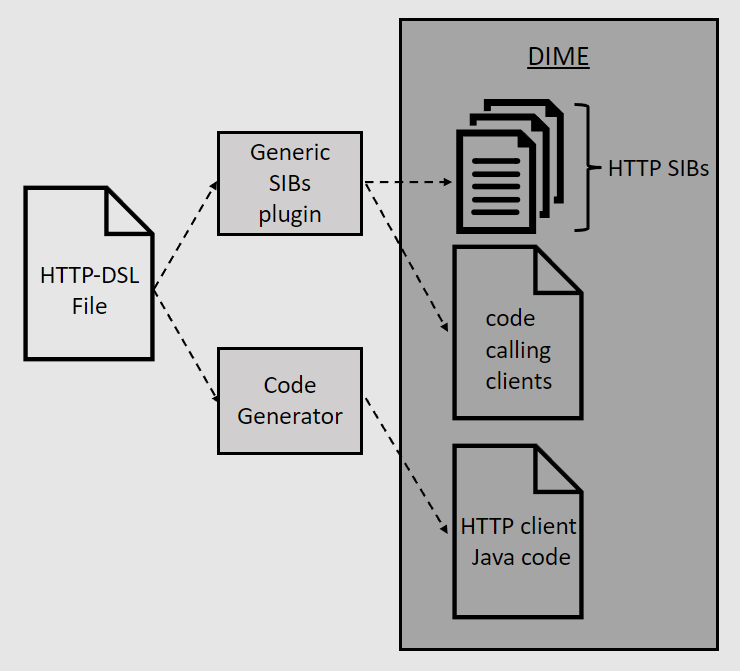}
		\caption{Overall structure of the solution.}
		\label{fig:mechanism}
	\end{center}
\end{figure}

\subsection{Aim of the Thesis}
The aim of this thesis is to facilitate a quick and easy access to web services within the web application modeling framework DIME.
To achieve this goal, we will take a deep look into web developers' needs, their  expectations regarding a solution, and develop a HTTP-DSL that provides all web developers with the ability to have a ready-to-use HTTP request in as little as five lines of code, as shown in Sect. \ref{subsec:efficiency}, and the press of a button.

\autoref{fig:mechanism} illustrates the structure of the solution, consisting of a DSL for HTTP, called HTTP-DSL, and the two code generating components: \textit{Generic SIBs plugin} and \textit{Code Generator}. 
The HTTP-DSL is a Xtext DSL that will allow the developers to quickly define HTTP requests. The requests are  automatically translated into a class structure (see \autoref{subsec:xtext}). 
The dotted lines denote fully automatic processes that will be triggered during the generation cycle of DIME. 
\begin{itemize}
    \item First, the \textit{Code Generator} generates a Maven~\cite{maven} project that contains the full codebase of the required HTTP clients. These are designed to send the previously defined HTTP requests (by using the HTTP requests' class structure). 
    \item Following this, the \textit{Generic SIBs plugin} automatically integrates these HTTP clients into the DIME application. This happens by providing the methods that will ultimately call the code previously generated by the \textit{Code Generator}. 
    \item Furthermore, the \textit{Generic SIBs plugin} will provide agile SIBs (\autoref{par:sibs}) that a web developer can simply drag and drop into models (\autoref{par:dimeLanguages}) to integrate HTTP requests into complex processes. 
    \end{itemize}
This structure allows for a completely automatic integration of HTTP requests into DIME, further extending the DIME language family.

\subsection{Structure of the Thesis}
This thesis starts with an analysis of expectations and requirements regarding such an integration of web services into DIME in \autoref{sec:Requirements}. The following section, \autoref{sec:Fundamentals}, will introduce the necessary background concepts and the technologies that will be used to implement the solution. The design and actual implementation will be discussed in \autoref{sec:Implementation}, including the initial design choices (\autoref{subsec:designSolution}), the DSL implementation (\autoref{subsec:DSL}), the code generator (\autoref{subsec:codegenerator}) and the integration into DIME (\autoref{subsec:semanticIntegratio}). In \autoref{sec:application} the implementation will be tested by providing two application examples of the solution. To see whether or not the project was successful, \autoref{sec:evaluation} will evaluate the solution with respect to efficiency, lean software development and the previously established requirements. Future prospects and further goals for the project will be examined in the concluding \autoref{sec:futureProspects}. \\
The full source code of the project is available at \url{https://gitlab.com/scce/http-lib} and the examples at \url{https://gitlab.com/brunosteffenuni/http-example}.
\newpage
\section{Requirements}
\label{sec:Requirements}

This section aims to find tangible requirements for the implementation of HTTP clients within DIME. We first address in \autoref{subsec:Userstories} the collected user stories; they are then analyzed in \autoref{subsec:USAnalysis}.  

\subsection{User Stories}
\label{subsec:Userstories}
One of the best tools to describe a desired product is to use user stories as they can help to define expectations with clarity. A user story is an informal, natural language description of features of a software system \cite{userstories1}. They are written from the perspective of an end user or user of a system, and Mike Cohn promotes their use as a compact form of requirements in projects \cite{userstories2}, especially in connection with XP approaches to software development. 
 Furthermore, the process of eliciting this collection of user stories helped changing perspective in this project from the initial mindset of “{\it what can I do}” to the mindset of “{\it what is needed/wanted}”.
 
To come up with the user stories, seven master students of Computer Science and five PhD students in Computer Science were interviewed. Some of these students have never worked with DIME, others have used it to build applications, and there was also a group of students who actively developed the framework itself.
These interviewees could roughly be grouped into three user roles, to cover all the necessary angles:

\begin{enumerate}
\item The {\bf (Web)app developer}, who uses DIME with pre-existing plugins to program an application.
\item The {\bf DIME developer}, who creates new plugins and DSLs for DIME, with the goal to improve and advance the framework itself.
\item The {\bf system administrator}, who integrates DIME applications into existing environments.
\end{enumerate}

With these three user roles in mind, the outcome of the interviews could be translated into 14 user stories. These user stories will serve as a guideline for the design choices in the implementation process and they will furthermore be used a posteriori to evaluate the success of the final product. They are as follows:

\begin{enumerate}[label=\textbf{U\arabic*}]
\item \label{us:u1}
As an app developer, I want to send simple HTTP-requests, to retrieve data from external services (e.g. REST).
\item \label{us:u2} 
As an app developer, I want to integrate HTTP-calls into complex workflows.
\item \label{us:u3} 
As an app developer, I only want to handle status codes that are relevant for my program.
\item \label{us:u4} 
As an app developer, I do not want to configure advanced settings (connection managers, authentication schemes, cookie management or proxy servers) for my ordinary use of HTTP APIs\footnote{An Application Programming Interface, in short API, describes a set of tools to interact with an external service}.
\item \label{us:u5} 
As an app developer, I want to be able to configure advanced settings (See above) whenever necessary.
\item \label{us:u6} 
As an app developer, I want to limit the points of failure during the development process.
\item \label{us:u7} 
As an app developer, I want to focus on the semantics of my program rather than the implementation on code level.
\item \label{us:u8} 
As an app developer, I want the model to be the single source of truth to avoid roundtrip problems (see \autoref{par:roundtrip} and \autoref{par:SSOT}).
\item \label{us:u9}
As an app developer, I need reliable and bug free code. 
\item \label{us:u10} 
As a DIME developer it is important to me that my code is reusable and that I do not have to “reinvent the wheel” for each reoccurring request.
\item \label{us:u11} 
As a DIME developer, it is important to me that preexisting code is functional even when the underlying code generator is updated or even changed to another programming language (see section \ref{par:roundtrip}).
\item \label{us:u12} 
As a DIME developer, I want the new HTTP client solution to be integrated natively into the DIME language family.
\item \label{us:u13}  
As a DIME developer, I want to be able to use the same tools for both frontend and backend.
\item \label{us:u14} 
As a system administrator, I want to allow easy configuration of the service integration based on the deployment environment, without touching the code, by injecting service configurations (e.g. host name) through standardized environment variables.

\end{enumerate}

\subsection{Analysis of User Stories}
\label{subsec:USAnalysis}

\begin{table}[ht]
	\begin{center}
 		\begin{tabular}{|c || c | c | c | c | c |}
			\toprule
			User Stories & Simplicity & Complex & Configuration & Reusability & Single Source  \\ 
			& & Workflows 
			& & & of Truth \\
 			\midrule
 			\midrule
 			\ref{us:u1} & X &  &  &  &      \\
 			\ref{us:u2} &  & X &  &  &      \\
  			\ref{us:u3} & X & X & X &  &     \\
   			\ref{us:u4} & X &  &  &  &      \\
 			\ref{us:u5} &  & X & X  &  &     \\
 			\ref{us:u6} &  &  &  &  &  X  \\
 			\ref{us:u7} & X &  &  &  &   \\
 			\ref{us:u8} &  &  &  &  &  X   \\
 			\ref{us:u9} &  &  &  &  &  X  \\
 			\ref{us:u10} & X &  &  & X  &        \\
 			\ref{us:u11} &  &  &  & X & X       \\
 			\ref{us:u12} &  &  &  & X  &      \\
 			\ref{us:u13} &  &  &  & X  & X     \\
 			\ref{us:u14} & X &  & X &   &    \\
			\bottomrule
		\end{tabular}
		\caption{Analysis of user stories gathered in \autoref{subsec:Userstories}.}
		\label{tab:grouping}
	\end{center}
\end{table}

A first analysis of these 14 unique user stories groups them  together to extract a few key objectives. Those objectives are important because they will emphasize where to focus the attention during design, implementation and evaluation, guiding the choice of the best fitting technologies. This grouping of the user stories is shown in \autoref{tab:grouping}. It identifies as key objectives simplicity, the handling of complex workflows, configuration capabilities, reusability and the single source of truth principle.  These key objectives will now be further discussed.

\paragraph{Simplicity.}
Looking at \ref{us:u1}, \ref{us:u3}, \ref{us:u4}, \ref{us:u7}, \ref{us:u10} and \ref{us:u14} one can see that the first main objective of the solution is \textbf{simplicity}. The users of the product do not want to invest a lot of effort to learn and understand how to use certain tools when they could simply program a HTTP client in their native programming language. Hence, it would make sense to create something that is trivial to understand for programmers and easy to learn for non programmers.

\paragraph{Complex Workflows.}
The next group, consisting of \ref{us:u2}, \ref{us:u3} and \ref{us:u5}, highlights that the integration of HTTP calls into \textbf{complex workflows} is absolutely necessary. The user story \ref{us:u3} is included in this group, because status codes in HTTP responses could potentially have various impacts depending on the workflows in which the initial HTTP request is integrated in. On a similar note, user story \ref{us:u5}, is included in this group because advanced settings are especially important when dealing with workflows where everything interlocks.\\
To achieve the goal of successful integration into complex workflows, the user needs to be able to chain HTTP requests together and process the received data in further requests. Additionally, it would make sense to make things modular, as it can be highly complicated to manage a design where several complex workflows are intertwined if there are no clear cut lines between them. 

\paragraph{Configuration.}
The third group of user stories consists of \ref{us:u3}, \ref{us:u5} and \ref{us:u14}, which all hint at the importance of \textbf{configuration}. In this context, configuration means that HTTP requests should be easy to extend with necessary details, and that the HTTP client in the background should be modifiable. These modifications could for example include alteration to the standard timeout, if the user knows that the accessed web service has a slow nature.\\
When talking about the first objective, simplicity, it was said that it is important to offer quick and easy "out of the box" HTTP calls. Such standard solutions can hopefully be widely used, however, they could be limiting at times, as they will not be able to cover every single kind of request the users might need. If possible, the solution should offer standard requests that the user can upgrade with custom configurations if necessary.

\paragraph{Reusability.}
The fourth objective is \textbf{Reusability}, which is mentioned in \ref{us:u10}, \ref{us:u11}, \ref{us:u12} and \ref{us:u13}. This aspect of the solution is mainly demanded by DIME developers, since they are the ones who possibly have to maintain the framework and web applications in the long run. \textit{Reusability} could either mean that the requests should be reusable within the same web application or that the user could even reuse the HTTP requests in upcoming projects down the road. 
Especially the user story \ref{us:u12} hints at the demand for easy reusability of the solution in DIME overall, meaning that this group in particular has many dimensions to it.

\paragraph{Single Source of Truth.}
The final cluster of user stories are stories \ref{us:u6}, \ref{us:u8}, \ref{us:u9}, \ref{us:u11} and \ref{us:u13}, which mention \textbf{single source of truth} and an attempt to have bug free code. These components all hint towards the use of one model or one specification file, in both cases a single one,  from which the code will be generated. This would limit the amount of bugs in the final code and allow for multiple coding artifacts to be generated from one specification, which addresses \ref{us:u14} directly.

\newpage
\section{Fundamentals}
\label{sec:Fundamentals}

This section describes and briefly illustrates the various necessary theoretical aspects that are required for this thesis. In \autoref{subsec:bgdefs} important definitions and concepts are discussed, which will be needed throughout the whole thesis. Then, \autoref{subsec:HTTP} explains what exactly is HTTP and its structure, followed by \autoref{subsec:xtext} which shines light on Xtext and its key features. The following \autoref{subsec:dime} introduces the main technology used in this thesis: DIME. Finally, \autoref{subsec: genericsibs} discusses Generic SIBs and their implementation, because this is a vital technology for the integration of DSLs into DIME.

\subsection{Background Definitions}
\label{subsec:bgdefs}
To help read and understand this thesis, relevant definitions of background concepts are gathered in this section.

\subsubsection{Domain Specific Language (DSL)}
\label{subsec:WhatIsDSL}
\textbf{Domain specific languages}, short DSLs, is a concept with very blurred boundaries \cite{Fowler10}. In fact, there are many programming languages where arguments could be made both ways. In his book "\textit{Domain Specific Languages}"  \cite{Fowler10},  on page 27, Martin Fowler defines DSLs as:
\begin{quote}
    "A computer programming language of limited expressiveness focused on a particular domain."
\end{quote}

\paragraph{Characteristics of a DSL.}
\label{par:DSLCharacteristics}
Fowler further explains that DSLs should have four distinct characteristics:
\begin{enumerate}
    \item As for all programming languages, the language needs to be instructed by a human and \textbf{executable for a computer}.
    \item The DSL must allow a fluent composition of the individual expressions allowing for a  "\textbf{language nature}".
    \item A DSL should be easy to use for a specific purpose, thus, it should have \textbf{limited expressiveness} to not allow too much of a leeway.
    \item Compared to a general purpose language~\footnote{General Purpose Languages are programming languages such as Java, C or C++ that can be used to program in a wide variety of domains.}, a DSL should be used for a particular aspect of a system, meaning it is \textbf{focused around a certain domain}.
\end{enumerate}
With these characteristics in mind it should be possible to identify whether a language is a DSL, and they certainly help when designing one. However, this still leaves us with an extensive spectrum of languages that fall under the DSL category.  

\paragraph{DSL Distinction.}
\label{par:DSLDistinction}
To distinguish between different kinds of DSLs and to narrow down their nature, Fowler specifies two categories of DSLs:
\begin{enumerate}
    \item \textbf{External DSLs.} An external DSL is a language that is completely separated in syntax from the core programming language of the system. This does not mean that the syntax of the DSL has to be unique, especially markup languages such as XML~\footnote{The Extensible Markup Language is a markup language used for the hierarchical representation of structured data in text files.} are very popular.
    \item \textbf{Internal DSLs.} Internal DSLs are languages that are derived from a particular application of general purpose languages. This could be achieved for example by allowing only a certain subset of the languages, to produce a new style of language that can only be used for a specific domain.
\end{enumerate}

\paragraph{Is HTML a DSL?}
 Confidently recognizing and categorizing DSLs is not an easy feat, as can be seen with the example of HTML\footnote{The Hypertext Markup Language is a markup language typically used to design web pages.}.
 On the one hand, HTML is designed for the web application domain, to design web pages for example. However, lately it is frequently used to create presentations, which is very well possible, since it could be used for designs of all kinds (especially if matched with CSS \footnote{Cascading Style Sheets is a language that is most often used to style HTML files.}).\\
When considering the characteristics of DSLs of \autoref{par:DSLCharacteristics}, it is clear that HTML is \textbf{executable by computers}, has a \textbf{language nature}, is limited in its \textbf{expressiveness} (HTML is by no means Turing complete) and is designed around the domain of web interfaces. With all these criteria fitting, HTML can certainly be regarded as a DSL.
In particular it could be further categorized as an \textbf{external DSL} since it has its own unique syntax.

\subsubsection{Single Source of Truth}
\label{par:SSOT}
The  \textbf{Single Source of Truth} paradigm (SSOT) is used in information systems to control the way models and data bases are structured. If implemented perfectly, every single piece of data can only be edited or altered in one place, meaning in one data schema. The object can then be referenced in others schemas, that are however read-only. The SSOT paradigm aims to eliminate errors such as duplicate entries or race conditions, in order to ensure that the data is valid, authentic and up-to-date.\\
This concept, even though it is taken from information systems, can be applied to a variety of fields. The idea that modifications can only be practiced in one single model, which can be then referenced or used to generate code, has the same advantages of enforcing coherence and consistency as in the data context from which it originates.

\subsubsection{Round-trip Engineering}
\label{par:roundtrip}
The round-trip engineering concept concerns techniques and efforts to synchronize multiple related software artifacts, such as models, code, or configuration files.
Round-trip Engineering is bidirectional, meaning that whenever an alteration to any of the related artifacts is taking place, all the others need to be updated accordingly.\\
The most common example is the problem of keeping consistency between a class diagram \cite{uml}, which is a Unified Modeling Language (UML) model, and the generated code stemming from it, such as class structures. In this case one has two artifacts: the class diagram and the source code. Two phenomena can occur when synchronizing these artifacts:
\begin{enumerate}
    \item Forward Engineering: A change of the class diagram triggers the generated source code, the class structure, to be updated accordingly.
    \item Reverse Engineering: A change in the class structure on code basis automatically triggers the class diagram to be updated.
\end{enumerate}
Keeping the consistency of multiple related artifacts, such as models and source code, can be very challenging. Modifications on the code level often can not be reflected in the diagram due to the higher level of abstraction in the model. Thus, a unidirectional approach to consistency is often the only viable option.

\subsubsection{eXtreme Programming}
\label{par:xp}
XP~\cite{BecAnd2004,XP} is a software development paradigm that uses collaborative techniques to put emphasis on customer requirements during the entire development process. There are countless practices that could be included in a broad understanding of XP. Since the main purpose of XP is to guarantee code quality under constantly changing customer requirements, a good modern example of XP adoption is CI/CD \cite{CICD}. \\
CI/CD stands for Continuous Integration and Continuous Deployment, a practice that enables developers to regularly show customers the latest version of their software projects deployed as a running system. To achieve this, tools such as CI/CD pipelines can be set up to automate the software life cycle steps, including testing, integration and deployment. This way, the code that is pushed by developers in their respective version control systems is quality-checked and deployed automatically, thus generating a fully functional version of the program, if the checks and tests were successful.\\
The main advantage of an XP practice is that customers can regularly see and evaluate the program during the development process. Customers can give frequent feedback and adapt their requirements for the software project when needed, to achieve a better end result.

\subsubsection{Extreme Model-Driven Development(XMDD)}
\label{par:xmdd}
XMDD \cite{Margaria2020} is a software development paradigm that combines the concepts of the previously discussed eXtreme programming (XP) paradigm and Model-Driven Software Development (MDSD) \cite{BrCaWi2012}. MDSD indicates a group of tools and techniques to generate source code from formal models. XMDD uses a formal approach too, and takes it further by demanding the program to be generated or at least portrayed in a single formal model. This model can consist of building blocks that contain software artifacts, however, its higher level of abstraction compared to only using source code ensures a better grasp over complex matters. Furthermore, these models are easier to interpret for customers or team members who may not come from a programming background, thus achieving without code the same goals that eXtreme programming is aiming at.

\subsubsection{Language Driven Engineering (LDE)}
\label{subsubsec:LDE}
The paradigm of Language-Driven Engineering (LDE)\cite{LDE}, {\it "is characterized by its unique support for division of labour on the basis of Domain-Specific Languages (DSLs) targeting different stakeholders"}. The core concept and mechanism used in LDE are different DSLs, tailored to the abilities and interests of the users. Also important is the use of model transformations and code generators that produce code from the DSL models. 
In the initial paper defining the LDE paradign~\cite{LDE}, DIME itself is introduced as a specific Mindset-Supporting Integrated Development Environment (mIDE) for web application development.

\subsubsection{Service Independent  Building Block (SIB)}
\label{par:sibs}
SIBs~\cite{ITUGlossary} are model-level components of larger scale applications that are (formally) well defined, connected to an implementation, and application-independent in the sense of being reusable in various contexts. The aim of the SIBs is the virtualization at the model level of the specific implementation of their body, and of the platform where they run. This is therefore a useful abstraction whenever one wishes to use functionalities in a heterogeneous technological landscape. The concept of SIBs stems originally from the telecommunication domain, where the International Telecommunication Union (ITU) introduced them to overcome the difference of proprietary implementations and execution environments for Intelligent Network services in telephony.  SIBs can contain a wide variety of services, from basic operations on a set of variables to complex procedures that affect the application as a whole. Regardless of the complexity of the service, it is contained in a simple interface, a building block, that can be integrated into further operations to reference the initial service.

\subsection{HTTP}
\label{subsec:HTTP}
The HyperText Transfer Protocol, short HTTP, is an application layer protocol widely used to load, store or alter files from web servers. These are most commonly HTML files that are displayed in a web browser.
HTTP is standardized by both the Internet Engineering Task Force (IETF) \cite{ietf} and the World Wide Web Consortium (W3C) \cite{w3}, and further upgrades surrounding the protocol are also implemented by the IETF. The standardization defines the components in a HTTP message and all the options that are made available in the protocol.

The most common version, HTTP 2.0, is used in 46.8\% \cite{HTTP1} of websites, however the subsequent version HTTP 3.0 is growing in popularity. Taken together, both versions are applied in more than 70\% \cite{HTTP1} of websites, making it the most widely used application layer protocol in the internet. 

In 2000 the HTTP was extended in RFC 2818 \cite{rfc2818} by encrypting the communication and authenticating the host server, resulting in the Secure HTTP protocol HTTPS (the appending 'S' standing for 'secure'). Here, the data is secured by encryption using the Secure Sockets Layer (SSL). The authentication is done via certificates that can be acquired from Certificate Authorities, which are trustworthy instances that sign digital certificates. The additional security layer compared to the HTTP mainly prevents man-in-the-middle attacks and provides internet users and web applications with a safe connection to trusted websites and web services.\\

\begin{figure}[h]
	\begin{center}
		\includegraphics[width=\textwidth]{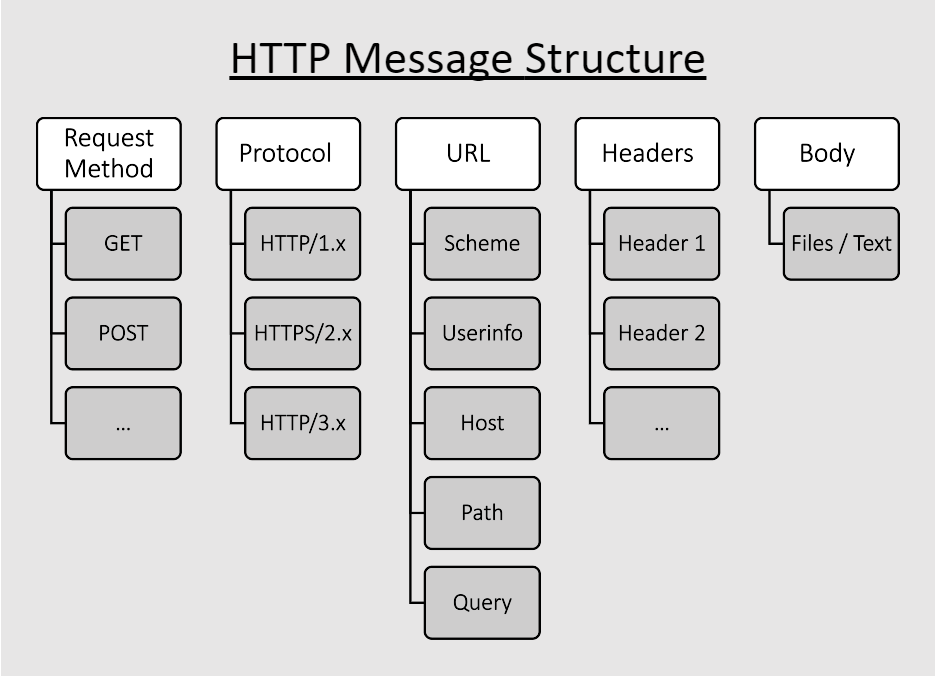}
		\caption{Components of a HTTP message.}
		\label{fig:http}
	\end{center}
\end{figure}
\begin{figure}[h]
	\begin{center}
		\includegraphics[width=\textwidth]{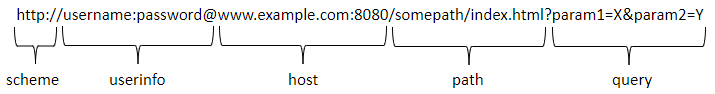}
		\caption{Components of a URL.}
		\label{fig:url}
	\end{center}
\end{figure}

As shown in Fig.~\ref{fig:http}, HTTP 2.0 messages consist of a request method, the protocol version, the uniform resource locator (URL), the headers and a message body. From a HTTP client's point of view, these message elements can be described as following. 
\begin{itemize}
    \item {\bf Request methods} define what kind of request the HTTP message is sending. They are categorized into nine different methods, the most commonly used ones being GET to retrieve a file, POST to send a file, UPDATE to alter a file and finally DELETE to delete a file. 
    \item The {\bf protocol version} serves the purpose to communicate to the receiver the version of HTTP the message is following. Since the specifications of HTTP/2 and HTTP/3 messages differ, this info is critical for the interpretation of the received message. 
    \item The {\bf URL}, illustrated in \autoref{fig:url}, is subdivided into multiple segments with unique properties. 
\begin{enumerate}
    \item The first segment, the \textbf{scheme}, is used to define which protocol is used in the current message. We are talking about URLs of HTTP messages, so the protocol will always by HTTP or HTTPS. The receiver could potentially use this information to help interpret the received message.
    \item The second segment is an optional \textbf{userinfo}. The sender can add credentials to get access to sources that may only be given to authorized users. However, since there are many more mechanisms for authentication, the userinfo segment is rarely used.
    \item The third segment, the \textbf{host}, defines the address of the host. This can be done in multiple different ways, either by specifying an IPv4 or IPv6 address, or by entering a fully qualified domain name (FQDN). The latter approach is used to give websites an address that is easier to interpret by humans, however, in the background every FQDN is mapped to one or more IPv4 or IPv6 addresses.
    \item The fourth segment, the \textbf{path}, is needed to specify the location of the desired file within the server. Oftentimes, especially when accessing file servers, the path can be directly translated to the file location from the root folder of the server.
    \item With the specified segments one could fetch any file stored on web servers that support HTTP, however there is one more segment, the \textbf{query}. It consists of parameters that can be used to give the server additional information that can be useful, for example for searching or filtering for the expected result.
\end{enumerate}
\item The next element of a HTTP message are the {\bf headers}, given as a list of key-value pairs. Headers are used to feed the server with additional information about the client or to specify the expected response. Some common headers are {\tt Accept} to specify the acceptable media type for the response, {\tt Cookie}, used to send the server a cookie, and {\tt Content-Type} to specify the type of content the message body is carrying. There are many more standard headers, and developers could define custom headers for a server to listen to.
\item 
The final element of a HTTP message is the {\bf message body}. It contains data and is only used for POST or PUT request methods, as these are the only methods for uploading data to a server. Since the data transported in the body could be of any data type, the {\tt Content-Type} header is especially important to use.
\end{itemize}

\subsection{Xtext}
\label{subsec:xtext}
Xtext can be described in multiple ways. One could view it as a set of tools that can be used to assist with Model Driven Software Development \cite{BrCaWi2012} (MDSD). MDSD describes techniques to produce running code from formal models, and Xtext fits into this approach perfectly.  

However, there is also a second angle from which one can describe Xtext.
Eclipse, the organization that owns and maintains Xtext, writes on their website that
\begin{quote}
"Xtext is a framework for development of programming languages and domain-specific languages. With Xtext you define your language using a powerful grammar language. As a result you get a full infrastructure, including parser, linker, typechecker, compiler as well as editing support[..]" \cite{Xtext}
\end{quote}
This excerpt outlines Xtext very well for users in terms of the key features it provides compared to a typical parser generator like ANTLR. Concretely, the user can define a language using a BNF-like grammar, and Xtext then generates for this language multiple useful tools:
\begin{enumerate}
    \item A parser, a linker, a typechecker and a compiler, which will be needed down the line when compiling the files of the new language.
    \item The EMF \cite{EMF} meta model of the given language. In brief, the EMF is a framework and code generation facility that is used to model meta models. The EMF meta model is the class model of the Abstract Syntax Tree (AST) that is generated from the language. This means that the user can generate fully functional classes of the different components in the specified language with ease.
    \item Tools to analyze and validate the AST.
    \item IDE concepts such as plugins, that can ultimately be used to turn the IntelliJ \cite{intellij} or Eclipse IDEs into fully functional IDEs for the specified language.
\end{enumerate}

Additionally, Xtext is supported on all editors and tools that support the language server protocol, like IntelliJ and Visual Studio Code \cite{vscode}. Xtext also has web editor support, meaning that it can be used in the web via servlets.

\subsection{DyWA Integrated Modeling Environment (DIME)}
\label{subsec:dime}
\paragraph{Introduction}
DIME (DyWA Integrated Modeling Environment) \cite{BFKLNN2016} is an integrated modeling environment that offers an array of graphic DSLs for web development. It belongs to the group of low-code application development tools and it has been used for industrial scale projects~\cite{DBLP:conf/ifm/MargariaS19,9568378,industrial3}. The main goal of this particular modeling environment is to offer DSLs that can be easily understood by domain experts (who might not be able to code), thus enabling them to program a web application on their own, ideally reusing and configuring existing building blocks that can be provided with the basic DIME tool or shared from other projects. This approach has the main advantage of eliminating many errors caused by misunderstandings between domain experts and programmers or web developers.\\
DIME follows the XMDD (eXtreme Model-Driven Design)~\cite{Margaria2020} and OTA (One Thing Approach)~\cite{Margaria2020} paradigms, meaning that the final code of the product is fully generated from the models, and not meant to be changed. Updates, bug fixes or alterations to the code are expected to be done by either changing the models or updating the code generators, since any alterations to generated code would be overwritten in the subsequent generation cycle.\\
The current version of DIME is based on the Eclipse Rich Client Platform (RCP)~\cite{RCP}  and has been developed in the CINCO SCCE Meta Tooling Suite~\cite{NaLyKS2017}. 
The frontend is realised solely using the Angular 2 framework \cite{angular} while the backend is written in Java, and the persistence layer is realized using PostgreSQL.
However, these infrastructural aspects of DIME are hardly noticeable for the user, because the framework offers a palette of DSLs that abstract from the actual implementation of the application. Specifically, it provides a number of modelling languages as well as a modelling environment. 

\paragraph{Modeling Languages.}
\label{par:dimeLanguages}
Firstly, the {\bf Data Language} is used to specify a domain model consisting of data types. At runtime, data objects of these types are created and managed by the web application. These objects are also referred to as "\textbf{domain objects}". These are the data objects used within both the \textbf{GUI Models}, that define the structure of the user interface, as well as the \textbf{ProcessModels}.\\
The {\bf Process Language} is designed for creating modular building blocks that can be cross-referenced to form a hierarchical structure that controls the business logic of the application.\\
To enable this, DIME offers {\bf Service Independent Building Blocks} (SIBs), discussed in \autoref{par:sibs}, that are used to link the services and models together through a uniform interface. 
In DIME, SIBs are represented by nodes within the models. These nodes can have:
\begin{enumerate}
    \item \textbf{Input Ports}, that are used to feed information into the SIB.
    \item \textbf{Branches}, that describe possible outcomes of the SIB (e.g., success and fail).
    \item \textbf{Output Ports}, that are attached to the branches to return data from the SIB.
\end{enumerate}
The integration of these SIBs within DIME has the main advantage that previously modeled processes or simple operations are packaged into atomic components which can then be used in further models.\\
In the last model language, called {\bf DIME Application Descriptor} (DAD) Language, the user configures the application as a whole. One can specify the landing and root pages of the application by linking the respective interaction models. In the DAD the user also has to declare all the relevant domain models (e.g., the data models) and relevant entities, such as user entities (used for the login).

\paragraph{Modeling Environment.}
\label{par:modelingEnb}
\begin{figure}[h]
	\begin{center}
		\includegraphics[scale=0.58]{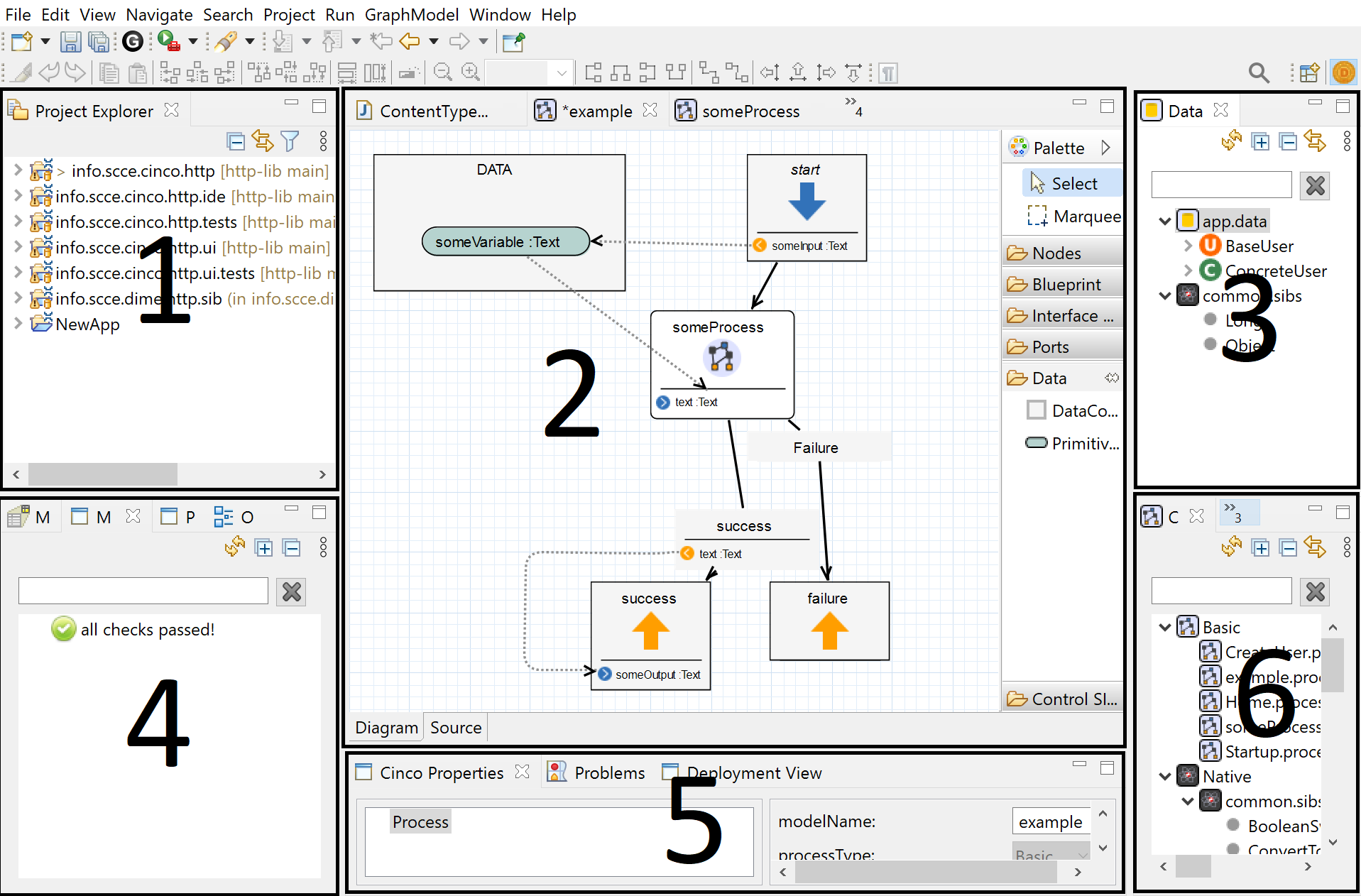}
		\caption{Screenshot of the DIME Modeling Environment.}
		\label{fig:modelview}
	\end{center}
\end{figure}

Figure~\ref{fig:modelview} shows the DIME modelling environment as it appears to a user, with its different zones and functionalities.
\begin{enumerate}
    \item \textbf{Project Explorer (1):} This is a typical project explorer as known from other eclipse applications such as the eclipse IDE.
    \item \textbf{Diagram Editor (2):} The Diagram Editor is the editor in which process SIBs, GUI SIBs or the DAD can be created by dragging and dropping items from the palette (seen on the right) or from the SIB palette in the Models View. These diagrams are control flow diagrams, that start with a start SIB (blue arrow) and finish with the end SIB (orange arrow). Furthermore, these diagrams typically contain a data context (box labeled \textit{data} in the top-left corner) in which local and global variables can be set and inserted into the components.
    \item \textbf{Data View (3):} The Data View is the view containing the data types that are defined in the data model, in files with the \textit{.data} file extension. These can be dragged and dropped into the models in the Diagram Editor.
    \item \textbf{Model Validation View (4):} This view contains the feedback from the model validators. The native DIME model validator checks for example whether every node in the control flow diagram has an outgoing edge to another node until the end SIB is reached.  
    \item \textbf{Properties View (5):} In the Properties View one can set various attributes of a component that was selected in the Diagram Editor. That way one can, for example, change the type of a variable or insert static data into components.
    \item \textbf{Models View (6):} The Models View contains the SIBs that are defined within the current environment. These SIBs can be Process model SIBs, GUI model SIBs, Native SIBs or Generic SIBs. Native SIBs are small SIBs for basic operations such as \textit{StringEquals}, where two Strings are compared. Generic SIBs are discussed next in \autoref{subsec: genericsibs}.
\end{enumerate}

\subsection{Generic SIBs Plugin}
\label{subsec: genericsibs}
The biggest difficulties in working with  DSLs often lie in their integration into existing programs or frameworks. Commonly, a manual integration of the DSL is needed, typically requiring knowledge of the architecture of the existing framework. This manual integration can lead to errors in the DSL or even errors in the preexisting framework. Consequently conventional integrations of DSLs are in general a significant temporal and monetary investment.

To combat the hurdle of integrating DSLs into the DIME framework we use instead a different mechanism: {\bf Generic SIBs}, supported by the Generic SIBs plugin. A first version of Generic SIBs stems from the Masters thesis of Jonathan Th\"one at the TU Dortmund \cite{Thoene}. While the related code has since been updated and upgraded, the core principles have not changed. 

\subsubsection{Principle of Generic SIBs}
\label{subsubsec:principlesGenerics}
In order to integrate an external DSL into DIME, fundamentally, one has to register an existing DSL to the Generic SIBs plugin. 
This works for different kinds of DSLs, such as  graphical CINCO-DSLs~\cite{Kopetz2014} and textual Xtext  DSLs (see \autoref{subsec:xtext}).
One chooses the kind of SIB that shall be generated, for example Process SIBs or GUI SIBs, and then implements the respective interfaces.
The Generic SIB plugin then takes components from the description files of the given DSL and generates the corresponding SIBs, which ultimately represent these components within DIME.
These SIBs can be used in DIME models just like any other SIB (e.g. process model SIBs or native SIBs). 

\paragraph{Example of Generic SIBs: a Math-DSL.}
\label{par:genericSibExample}
  This paragraph entails an example, taken from the Masters thesis of Jonathan Th\"one~\cite{Thoene}, to illustrate in a simple fashion the functionality of Generic SIBs.\\
  The Math-DSL in this example is a graphical DSL used to describe basic calculation processes. This DSL was created using a framework called CINCO \cite{NaLyKS2017}, however, a detailed description of the CINCO framework is not necessary to understand the example and would exceed the scope of this thesis. The Math-DSL has a graph-like structure and contains four kinds of nodes:
  \begin{enumerate}
      \item \textbf{Start Node:} It indicates the beginning of the model and it must contain at least one variable node.
      \item \textbf{End Node:} It indicates the termination point for the execution of the model, thus it has only incoming edges.
      \item \textbf{Operations Node:} The Math-DSL defines one of these nodes for each of the basic mathematical operations: addition, subtraction, division, multiplication.\\ Each operation node has two incoming edges with numerical values and one outgoing edge as a numerical result. 
      \item \textbf{Variable Node:} It indicates a variable, that is implicitly of type real.
  \end{enumerate}
  In addition, the DSL contains two kinds of edges:
  \begin{enumerate}
      \item \textbf{Data Edge:} This edge type describes the data path for the operations, and shows how variables and results are used for further operations. In the model the data edge is represented by a dotted arrow.
      \item \textbf{Control Edge:} This edge denoted the control flow, i.e., the procedure of operations in a given model. In the model the Control Edge is represented by a solid arrow.
  \end{enumerate}
  
\begin{figure}[h]
	\begin{center}
		\includegraphics[scale=0.5]{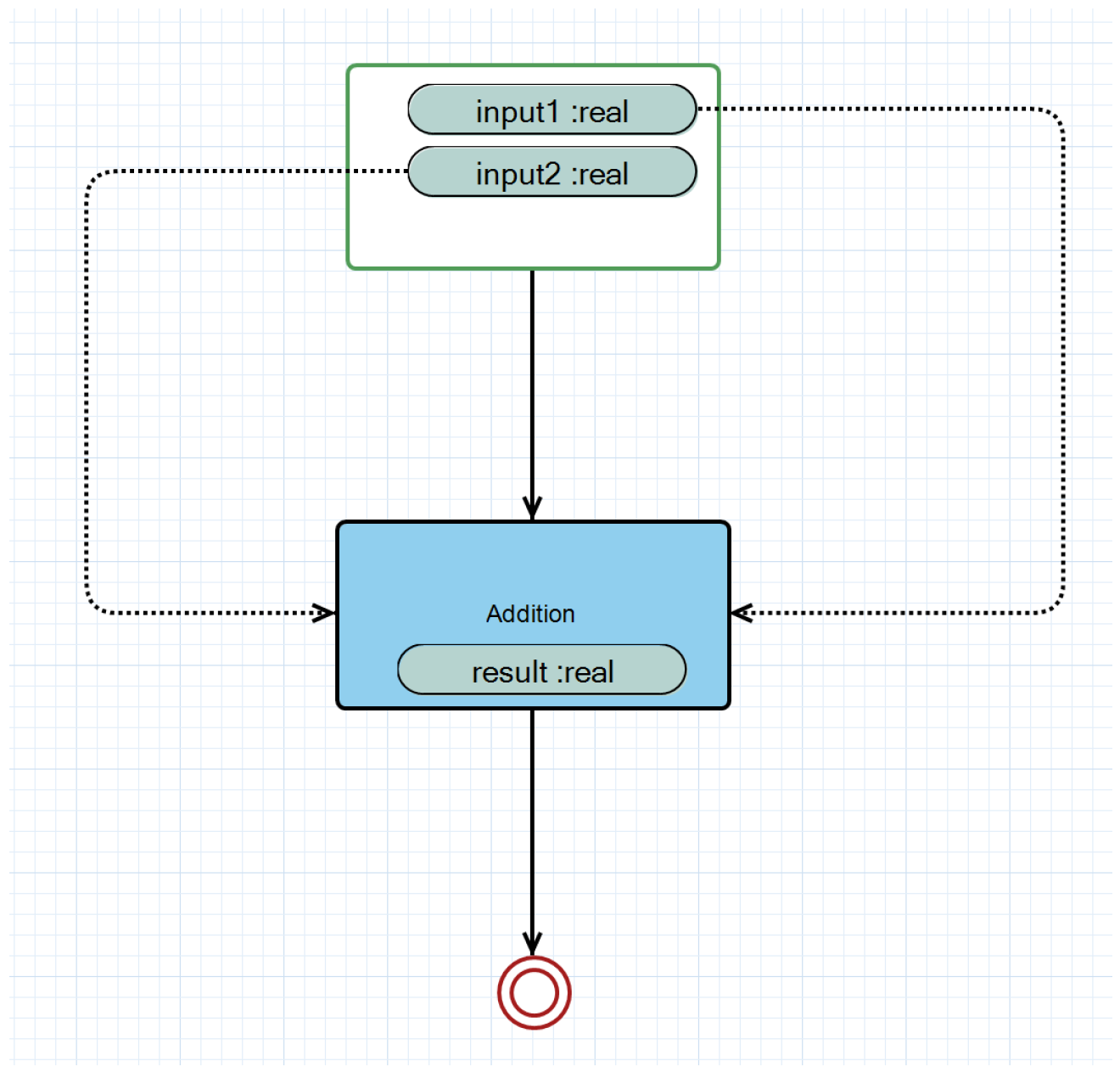}
		\caption{Math-DSL model of the addition operation (see \cite{Thoene} fig. 4.1). }
		\label{fig:genericSibDSL}
	\end{center}
\end{figure}

The first step for an integration is to model a mathematical procedure in the Math-DSL to obtain a DSL component that can later be translated into a DIME SIB. 
In this example, shown in~\autoref{fig:genericSibDSL}, the  component is simply the mathematical procedure of performing the \textit{addition} operation on two real numbers. This model contains a \textbf{Start Node} that holds two variables called \textit{input1} and \textit{input2}, receiving the input data. Their sum will be calculated by propagating the data to the {\tt Addition} \textbf{Operations Node} representing the addition computation, which happens by using data edges. The flow of the calculation is defined through the control edges from the \textbf{Start Node} to the \textbf{Operation Node} and the one ending in the \textbf{End Node}.\\
The other operations are defined analogously.

\begin{figure}[h]
	\begin{center}
		\includegraphics[scale=0.65]{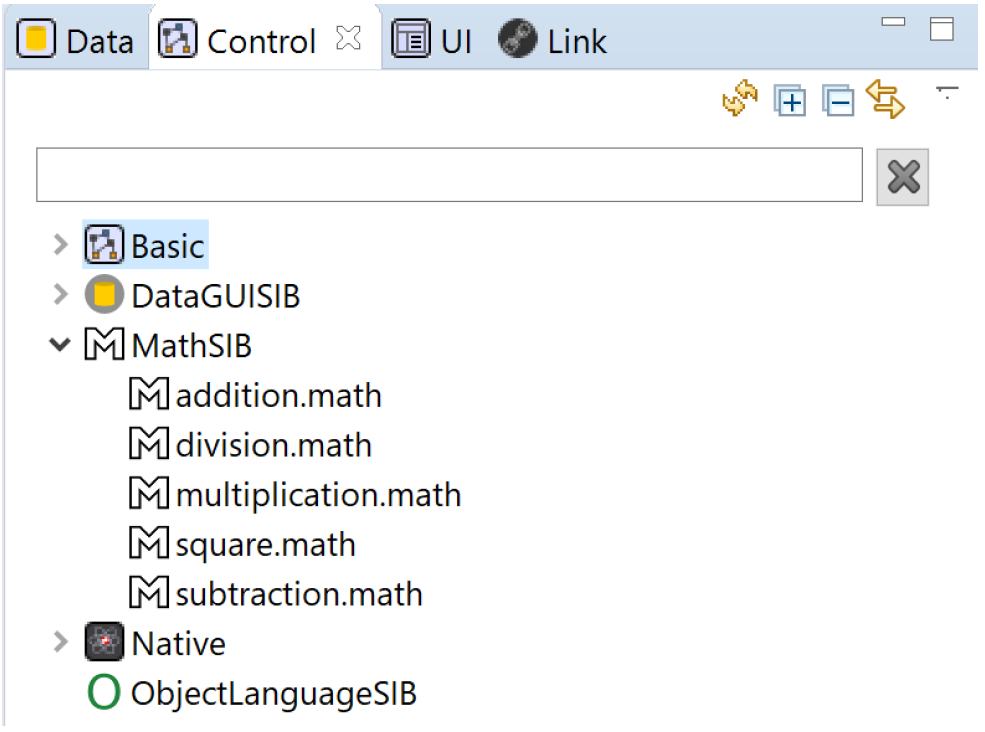}
		\caption{Models View of DIME application after saving the Math-DSL file (see \cite{Thoene} fig. 4.2).}
		\label{fig:genericSibMV}
	\end{center}
\end{figure}

Once the DSL definition is done, the user can save the model and if set up correctly, the Generic SIBs plugin (see~\autoref{subsubsec:howToGenerics}) creates the SIB that is then available in the \textbf{Models View} described in~\ref{par:modelingEnb}. This can be seen in~\autoref{fig:genericSibMV}, where the {\sf addition} SIB is now available next to the other mathematical operation SIBs in the DSL {\sf MathSIB}.

\begin{figure}[h]
	\begin{center}
		\includegraphics[scale=0.65]{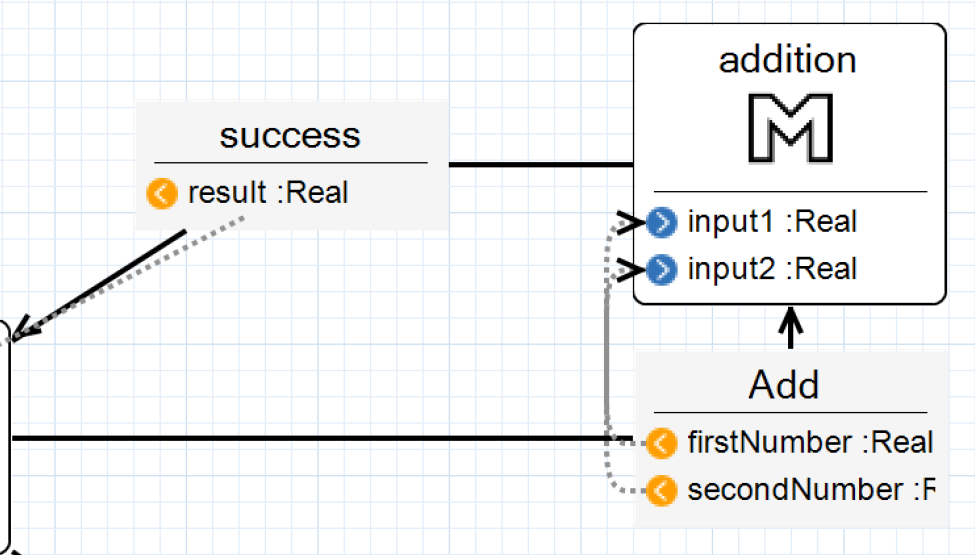}
		\caption{Example of use: Integration of the generated SIB into a DIME Process SIB (see \cite{Thoene} fig. 4.3). }
		\label{fig:genericSibSIB}
	\end{center}
\end{figure}

At this point, the design and integration are concluded, and the SIBs are ready to be used in DIME.\\
At use time, the last step is to take the generated SIB and to include it in a DIME \textbf{Process Model}, as described in Sect.~\ref{par:dimeLanguages}. This is illustrated in~\autoref{fig:genericSibSIB}\footnote{To simplify the image, only an excerpt was taken from the original example in~\cite{Thoene}}. 
This image clearly illustrates how the DSL variables seen in \autoref{fig:genericSibDSL}, called \textit{input1} and \textit{input2}, are automatically transformed into \textbf{input ports} of the SIB in the \textbf{Process Language} context.

During the deployment phase of the DIME application, the code generator of the Math-DSL generates the code that computes the defined calculations. Simultaneously, the Generic SIBs plugin automatically links that code to the functionalities of all the relying SIBs in the process models. This way the user does not have to do any further manual integration of the Math-DSL into the DIME application.

\subsubsection{How to Implement Generic SIBs}
\label{subsubsec:howToGenerics} 
To make use of generic SIBs, DIME developers have to implement three interfaces, namely \classname{I\-Model\-trafo\-Supporter}, \classname{IGenericSIBBuilder} and \classname{IProcessSIBGenerator}, this last one being specific for process SIBs. 

\begin{itemize}
    \item The \classname{IModeltrafoSupporter} interface has three distinct purposes. Its first purpose is to register implementations of the other two interfaces. The second purpose is to declare which files should be scanned to generate SIBs and how to scan them. The final purpose is simply to assign an icon that the SIBs will later on use.
   
    \item The  \classname{IGenericSibBuilder} interface is used to model the SIBs. Since process SIBs can have both inputs and outputs, the \classname{IGenericSibBuilder} interface offers the methods \classname{getInputPorts} and \classname{getOutputBranches} to model exactly those. Furthermore one can set the name and the label of the generated SIBs via the methods \classname{getLabel} and \classname{getName}.
    
    \item 
    The \classname{IProcessSIBGenerator} interface is the most complex one as its purpose is to provide the generation of the code that calls the DSL-generated code whenever the SIB is used. This was represented by the \textit{code calling clients} artifact in \autoref{fig:mechanism}. It works by using three methods: 
    
    \begin{itemize}
        \item The method \classname{getMethodCallSignature} links the code that is generated by the Generic SIBs plugin to the DIME application code. 
        \item The second method, \classname{generateContent}, has the actual templates for the code generation.
        \item The third and final method, \classname{getResultTypeName}, gives the DIME application the type of the expected result.  
    \end{itemize}
   If no further processing of the result from the DSL generated code is done, the result type is identical to the result of the method from the DSL generated code.
\end{itemize}

Once these three interfaces are implemented, Generic SIBs offer the service to create process SIBs fully automatically when the DSL configuration file is saved. These SIBs can then be used in processes like any other SIB would, without the need of touching the code of the actual framework.

\newpage
\section{Design and Implementation}
\label{sec:Implementation}
The purpose of this section is to explain and discuss the design of the solution and the actual implementation of it. This is why this section is divided into two parts, beginning with the design component in \autoref{subsec:designSolution} followed by the actual implementation in the subsequent sections.

\subsection{Design of the Solution}
\label{subsec:designSolution}
Before the implementation, it is important to discuss how a possible solution can look like.
Section \ref{subsubsec:IRC}  discusses why certain technologies are used within this project with regards to the requirements of \autoref{subsec:USAnalysis}. Then \autoref{subsubsec:concreteDesign} presents the concrete design whilst introducing the implementation.

\subsubsection{Identifying Required Components}
\label{subsubsec:IRC}
The solution must be easy to understand, reliable, reusable and easy to integrate into existing workflows. The following paragraphs discuss why certain design choices were made and how they reflect the requirements made in \autoref{subsec:USAnalysis}.

\paragraph{Using DIME.}
\label{par:usingDime}
DIME has a few characteristics that fit the criteria very well. DIME follows the XMDD concept, introduced in \autoref{par:xmdd}, which addresses the expectations of \ref{us:u11}. The whole code base of DIME applications is generated from models, meaning that any round trip concern (as defined in Section~\ref{par:roundtrip}) is dealt with.\\
Furthermore, DIME makes use of SIBs (see \autoref{par:sibs}) which are inherently \textbf{reusable} building blocks. Thus anything that is modeled within DIME can be processed and integrated in further models.

\paragraph{Using a DSL.}
DSLs fit the criteria for this project on multiple ends. 
\begin{enumerate}
    \item The nature of DSLs is that they are meant to be easy to understand, so that domain experts can either use them by themselves or at least interpret the code. Web developers could be considered to be domain experts in the field of HTTP, as they are likely to come in touch with this protocol on many occasions throughout their careers. This means that for the typical DIME user, who wants to program a web application, a HTTP-DSL should be rather \textbf{simple} to use. For DIME users who are unfamiliar with web development, the higher abstraction level that a HTTP-DSL would offer compared to Java code should also simplify the interpretation of the code.
    \item The DSL should offer a standard solution that the user should be able to use with just a few lines of code. With this particular feature, a DIME user with no previous web development experience should be able to create HTTP requests with minimal knowledge. Additionally, this feature would guarantee that DIME users would have to \textbf{spend minimal time} to create elemental HTTP requests.
    \item The DSL should suite the expectations regarding \textbf{configurations}. Consequently, the user should be able to append desired configurations or customizations to the DSL code. This step should be completely optional, to keep a good balance towards the simplicity aspects as well.
    \item When creating a new DIME application, a user could simply take HTTP-DSL description files, meaning files in which DSL code is written, from previous projects and simply copy them into the newly created application. This way the HTTP requests could be \textbf{reused} not only within the same application, but even in further upcoming DIME applications.
    \item The introduction of a new DSL in DIME suits the LDE approach that DIME follows since its release. The idea of adding additional DSLs that tackle new problems within DIME fits its nature betterthan creating a custom Java HTTP client library that is referenced by native SIBs within DIME.
\end{enumerate}

There are multiple ways to design a DSL for the purpose of configuring HTTP requests. One possibility is a graphical DSL to drag and drop different components of the HTTP into some sort of HTTP objects. Such a HTTP object could look like the message structure shown in \autoref{fig:http}, which could certainly help users without experience in web development. However, the broad user base of DIME consists of web developers or users who have at least some programming experience. Hence, it makes sense to use a textual DSL. It certainly speeds up the process of defining the HTTP requests if used correctly, and it lightens the workload on the working machine. For textual DSLs, Xtext is a very good option. As discussed in \autoref{subsec:xtext}, it automatically generates the EMF meta model, providing a complete set of classes that represent the DSL.

\paragraph{Generating Code.}
\label{par:generatingCode}
Code generation fits in perfectly, as it realizes vital requirements for this project. The most obvious feature is that the previously mentioned \textit{Single Source of Truth} (discussed in \autoref{par:SSOT}) paradigm is systematically implemented. This means that all kinds of changes to the program are made on model or DSL level, instead of altering the actual Java code.\\
On the one hand this choice prevents syntactical errors in the Java code, on the other hand it puts the focus solely on the DSL. This means that special attention is paid towards \ref{us:u10}, because users will not have to think about the code level implementation of a HTTP client, but only about its configuration.\\
The code generator can use the EMF meta model that represents the HTTP-DSL description file and generate all the required code bits and pieces. The generated code is then integrated into the DIME application structure in order to have it ready and available during runtime.

\paragraph{Using the Generic SIBs Plugin.}
\label{par:designGenericSIBs}
In the context of DIME, including a tool into workflows can roughly be translated as “make it possible to include the tool into the Process Language” (section \ref{par:dimeLanguages}). With the HTTP-DSL and the code generator, the HTTP client code is included the project in form of Java files. However, unless there is also some form of integration within DIME models, the application will not know how to use that generated code.\\
The \textbf{Generic SIBs Plugin}, further discussed in \autoref{subsec: genericsibs}, closes this gap. It takes the DSL code and generates a SIBs and further needed code out of it. These SIBs, like SIBs within DIME in general, can then be used and integrated as nodes within DIME models.

\subsubsection{Concrete Design of the Solution}
\label{subsubsec:concreteDesign}
The overall mechanism of the solution is shown in \autoref{fig:mechanism}. 
Going from left to right, one can see that initially a \textbf{HTTP-DSL file} is required. This file defines the HTTP requests, holding the information that will be translated into a objects by xtext and used by the \textbf{Code Generator} and the \textbf{Generic SIBs Plugin} to generate all the HTTP request-related code within the DIME application. This code consists of:
\begin{enumerate}
\item the \textit{HTTP SIBs} code, that defines how the SIBs can be used within the DIME ProcessModels discussed in \autoref{par:dimeLanguages},
\item The \textit{HTTP client Java code}, that contains all the things necessary to ultimately send the HTTP requests that are defined in the HTTP-DSL.
\item  the \textit{code calling the clients} is called by the SIBs and used to access and call the classes and methods that are used to send the HTTP requests.

\end{enumerate}
The implementation will be split into three steps to produce the final product: the \textbf{HTTP-DSL}, the \textbf{Code Generator} and the \textbf{Generic SIBs Plugin}, that are elaborated in the coming sections. Specifically,  \autoref{subsec:DSL} explains how the \textbf{HTTP-DSL} is implemented and designed to suit this project. Then \autoref{subsec:codegenerator} covers the entire process of producing running Java code from a HTTP-DSL specification file. Finally, \autoref{subsec:semanticIntegratio} details how these components are integrated into DIME, to guarantee easy access to the DSL by web application developers.

\subsection{The HTTP-DSL}
\label{subsec:DSL}
Since the domain specific goal of the required DSL is to ease sending HTTP messages, the implementation is focused along the lines of the structure of HTTP messages previously introduced  in \autoref{subsec:HTTP}. Hence, the objects used in this DSL are named after the corresponding parts of a HTTP message. This naming decision is particularly important for wide usability because most developers have at least minor knowledge about the protocol. As seen in the first analysis of the user stories, simplicity and usability are very important. It is therefore more valuable in practice to create a DSL that does not need a complicated tutorial, one that is intuitive to use.\\
When finished, a HTTP request specification should look like the code depicted in \autoref{fig:dslEx}. Intuitively, it goes from giving a name to the message to setting the HTTP components such as URL, request type and parameters for the query. A deeper look into this specification is taken at the end of this section.\\

\begin{lstlisting}[
label={fig:dslEx},
style=HTTP,
caption={Example of a user-created description file in the HTTP-DSL.}
]
http{
	name "WeatherLocation"
	url server http://dataservice.accuweather.com
		path locations/v1/cities/search
	type GET
	param "apikey" : input $apiKeyParam
	param "q" : input $city
	param "language" : "en-US"
}
\end{lstlisting}

In the following, we describe the handling of a message's variables, essential fields and optional fields, and then illustrate how this comes together by means of a usage example. 
The full Xtext codebase for the HTTP-DSL can be found in \autoref{appendix:dslCode}.

\subsubsection{Variables}
\label{subsubsec:variables}
Since variables can be used for most of the fields in the HTTP-DSL, their implementation is discussed first. The use of variables in the DSL is essential because many expectations derived from the user stories heavily depend on transferring data.
To make \ref{us:u2} possible, concerning integration of HTTP calls into complex workflows, it is necessary to get data from one SIB to another, otherwise a complex workflow would not be possible. A second example for the requirement of variables is \ref{us:u14}, addressing easy configuration of the service integration, which specifically mentions environment variables. The mechanism for handling environment variables is different from the one for data variables, where the data is entered either by the developer or via the data flow in the process model of a DIME application. Furthermore, the code generator will have to distinguish between the kinds of variables, since each kind of variable will be implemented differently in the generated Java code. However, both versions of variables should be usable in the same inputs throughout the DSL, even if at certain points the one makes practically more sense than the other.\\

Accordingly, we introduce in the DSL a parent class \classname{AbstractVariable}, with two possible implementations:   \classname{InputVariable} and \classname{EnvironmentVariable}.

The \classname{InputVariable} is used to represent data coming from DIME via SIB input ports. These ports can be static, and thus filled while modelling, or the ports are primitive, which means that they are filled as the dataflow reaches a certain point during runtime.\\
The {\bf syntax for input variables} consists of the leading keyword \textit{input} followed by a '\$' sign as a prefix to the concatenated variable name (e.g., \textit{input \$url)}.

Instances of the second type of variables, \classname{EnvironmentVariable}, can only be entered at runtime, as normally done by operators. These variables are particularly useful when the design of the API is clear, but the actual endpoint is entered at runtime (often done when testing, e.g., with localhost:8080). \\
The {\bf syntax for environment  variables} is the leading keyword \textit{environment}, followed by a string of capitalized words connected by '\_' underscore signs (e.g., \textit{environment SERVER\_URL}). \\

The use of an abstract class eases potential future extension. This could happen for example for refinement, in case the need arises to distinguish between different environment types, or different data inputs, or if a completely new third kind of variables is introduced in new protocol versions. 
Variables are used in both the essential and optional HTTP message fields, that will be described next. \\

\begin{lstlisting}[
label={fig:dslImpl},
style=xtext,
caption={Xtext definition of the core structure of the HTTP-DSL.}
]
HttpFile: {HttpFile}
	messages += HttpMessage+
;

HttpMessage: 
	'http' '{'
		'name' name = STRING
		'url' url = AbstractUrl
		'type' requestMethod = RequestMethod
		query += Param*
		headers += Header*
		body = Body?
		returnValue = ReturnValue?
		customization = Customization?
		'}'
;
\end{lstlisting}

\subsubsection{Essential Fields}
\label{subsubsec:essentialFields}
As shown in \autoref{fig:dslImpl} line 1, the description files of the DSL include a list of at least one \classname{HttpMessage} object. This way it is possible for each description file to contain multiple HTTP messages, and this in turn can help structuring the code. Compared to a design where the files contain a single HTTP message, the user can now group the HTTP messages based on semantic groups, such as by type of messages or services that need to be accessed. 

\classname{HttpMessage} objects have three mandatory fields. The  \textbf{name} field gives the generated SIB a unique name. The other two fields, \textbf{url} and \textbf{type}, are necessary since without a valid URL and a request method one can not specify a message that follows the HTTP standards.

\paragraph{URL.}
The \textbf{url} field, as seen in \autoref{fig:dslImpl} line 5, holds an instantiation of the \classname{AbstractUrl} class. This class has two fields, \textbf{server} and \textbf{path}, where \textbf{path} holds the URL path and \textbf{server} holds all the preceding elements of a URL, namely scheme, userinfo and host, as described in section \ref{subsec:HTTP}.
Both fields can be initialized either by \classname{AbstractVariable} objects or by manually typing them into the specification file. The latter approach does not allow the fields to be filled dynamically during runtime, however it can be argued to be safer. This is due to the fact that the DSL can only validate if the given URL and path are syntactically correct, if they are entered in the specification file.\\
The implementation of this validation feature is done according to the IETF RFC 1738 section 5 \cite{RFC1738}, the RFC designed by the IETF to specify URLs. Since the  original IETF specification  provides a "{\it sort of BNF}", as the IETF calls it, one could translate the given rules into Xtext rules. 

The translated BNF can be seen in \autoref{fig:urlBNF}. Two details in the translated BNF differ from the RFC 1738 specifications.
\begin{itemize}
    \item The RFC 1738 BNF allowed a top level domain to have a single character. However, further research revealed that this should not be possible since the IETF itself wrote on their website \cite{TOPDOM} that a top level domain has a minimum of two and a maximum of 63 characters. The DSL now therefore supports the second option.
    \item The specifications RFC 1738 did not take IPv6 addresses into account. They were defined only later, in the RFC 1883~\cite{RFC1883}, and must now be supported too. Thus, the translated BNF allows IPv6 addresses which are syntactically correct with respect to the IPv6 specifications published in RFC 5854 \cite{RFC5854}.
\end{itemize}

\begin{lstlisting}[
label={fig:urlBNF},
style=xtext,
caption={Xtext definition of the URL.}
]
terminal fragment DIGIT:('0'..'9');
@Override
terminal INT returns ecore::EInt: DIGIT+ ; 
terminal fragment UCHAR: (ALPHADIGIT|"$" | "-" | "_" | "." | "+"| "!" | "*" | "'" | "(" | ")" | ",");
terminal fragment ALPHADIGIT: ALPHA | DIGIT;
terminal fragment ALPHAUP:('A'..'Z');
terminal fragment ALPHA: ('a'..'z')|ALPHAUP;

terminal SERVER:(PROTOCOL)? (CREDENTIAL ':' CREDENTIAL '@')? HOST ;
terminal fragment PROTOCOL: 'http''s'? ':' '//';
terminal fragment HOST: (HOSTNAME | HOSTIPV4 | HOSTIPV6) (':' INT)?;
terminal fragment HOSTNAME: (DOMAINLABEL '.')* TOPLABEL;
terminal fragment DOMAINLABEL: ALPHADIGIT | (ALPHADIGIT (ALPHADIGIT | "-")* ALPHADIGIT);
terminal fragment TOPLABEL: ALPHA (ALPHADIGIT | '-')* ALPHADIGIT;
terminal fragment HOSTIPV4: INT '.' INT '.' INT '.' INT;
terminal fragment HOSTIPV6:'[' HEXPART (':'HOSTIPV4)?']';
terminal fragment HEXPART: HEXSEQ | (HEXSEQ '::' (HEXSEQ)? )|('::'(HEXSEQ)?);
terminal fragment HEXSEQ: HEX+ (':' HEX+)*;
terminal fragment HEX: ('0'..'9')|('a'..'f')|('A'..'F');
terminal fragment CREDENTIAL:(UCHAR | ";" | "?" | "&" | "=")+;

terminal PATH: SEGMENT ('/' SEGMENT)*; 
terminal fragment SEGMENT:(UCHAR |";" | ":" | "@" | "&" | "=" )+; 
\end{lstlisting}

Accordingly, the DSL checks that the server, resp. path, are compatible with the format specified in \autoref{fig:urlBNF} lines 9-20, resp. 22-23.

\paragraph{Request Method.}
The \textbf{requestMethod} field, shown in \autoref{fig:dslImpl} line 10, holds a \classname{RequestMethod} enum field that allows users to choose from ‘GET’, ‘POST’, ‘PUT’ and ‘DELETE’.   Out of the 9 foreseen options by the HTTP standard, these are the four request methods implemented at the time of publication. An enum type, in contrast to a String, is superior for this scenario since it prevents users from provoking syntactical errors through typos.

\subsubsection{Optional Fields}
\label{subsubsec:nonessentialFields}
The non essential fields of a \classname{HttpMessage} object are \textbf{query}, \textbf{headers}, \textbf{body}, \textbf{returnValue} and \textbf{customization}.

\paragraph{Query.}
The \textbf{query} is defined in \autoref{fig:dslImpl} line 11 as a list that consists of at least one instance of a \classname{Parameter} object. Each parameter has a key-value pair, which can be defined in the description file either by entering a string or by referring to an \classname{AbstractVariable} object.

\paragraph{Headers.}
The headers field, defined in \autoref{fig:dslImpl} line 12, is a list containing at least one instance of the \classname{Header} object which, like the parameters, consists of a key-value pair. The \classname{Header} class is designed almost identically to the \classname{Parameter} class, with one key difference. Parameters in a HTTP message have no conventions for naming the key-attributes, however headers do. Consequently, the \classname{Header} class is designed differently. The \textbf{key} field in the \classname{Header} class, shown in \autoref{fig:headers} line 67, can be entered either by referring to an \classname{AbstractVariable} object, or by using an \classname{AbstractHeaderKey}.\\

\begin{lstlisting}[
label={fig:headers},
style=xtext,
caption={Implementation of header field in the DSL.}
]
Header: 'header' 
	(key = AbstractHeaderKey | key=AbstractVariable) ':' 
	(value = HeaderValue | value=AbstractVariable);
AbstractHeaderKey:(HeaderKey | CustomHeaderKey);
HeaderKey: value=HEADERKEYSTRING;
CustomHeaderKey: value=STRING;
HeaderValue: value=STRING;
enum HEADERKEYSTRING: AGE="Age"|ALLOW='Allow'|...
\end{lstlisting}

The \classname{AbstractHeaderKey}, in \autoref{fig:headers} line 4, is the abstract parent class of \classname{HeaderKey} and \classname{CustomHeaderKey}. Its first implementation alternative, \classname{HeaderKey}, is an enum containing the complete list of  header keys conventionally used in HTTP. It should be used by the user in any possible occasion to avoid errors. Its second implementation, \classname{CustomHeaderKey}, containing a simple string, is used to enter header key names that are either server specific or uncommon.  Users could potentially make use of the \classname{CustomHeaderKey} also when using the standard headers, however, this can be riskier considering that a typo could turn the header invalid.

\paragraph{Body.}
The \textbf{body} field, defined in \autoref{fig:urlBNF} line 13, is a \classname{Body} object containing two fields for the specification of the payload (\textbf{contentType}  and \textbf{entityType}) and one field for the actual \textbf{payload}. \\
The \textbf{contentType} field refers to an instance of either an \classname{AbstractVariable} object, a \classname{ContentType} object, or a \classname{CustomContentType} object, and it is used to automatically set the “content-type” header during code generation. This mechanism of classes is analogue to the classes previously used for the header key.
The user can either use a variable, insert a content type via a string or the user can choose from an array of commonly used media types such as “text/plain” or “image/jpeg”. Similar to the benefits of having a list of commonly used header keys we saw before, the benefit of having such a list for content types potentially decreases the errors in the code.\\
The \classname{EntityType} object is the only configuration in the DSL that does not derive from the domain, but from the underlying http client library used during code generation. The library provides four alternative ways to insert the payload in the HTTP message:
\begin{enumerate} 
\item The first option is a \classname{StringEntity} which is used for most HTTP messages, for example when sending a JSON. 
\item The second option is a “\classname{FileEntity}” which would include most file formats, however the user could also use the
\item  “\classname{InputStreamEntity}” or
\item “\classname{ByteArrayEntity}” to send Files.
\end{enumerate} 
The latter options have the effect that the HTTP messages are not self contained, meaning that the message does not contain the whole payload, but it can open a stream to access it. Since these four options are those available in the Apache HTTP client library, the user is presented this lists and has only to choose the appropriate one.\\
\noindent The final field of the \textbf{body} is the \textbf{payload}. It can be set by using a string or a variable, however especially when transmitting files or streams it is highly unlikely that the user should use a static string.

\paragraph{Return Value.}
The \textbf{returnValue} field, seen in \autoref{fig:dslImpl} line 14, is used to define the expected response of the HTTP message and how it should be encapsulated. Every \classname{ReturnValue} object has two fields: \textbf{expectedType} and \textbf{returnForm}. \\
The \textbf{expectedType} field can hold a \classname{ContentType} object, a \classname{CustomContentType} object or an \classname{AbstractVariable} object. It works in an  analogue way to the \textbf{contentType} field in the body.\\
The \textbf{returnForm} field on the other hand is unique and holds an enum value which has a choice of two values. The user can either get the entire response message, which is a native java \classname{HttpResponse} object, or specify to get only the payload of that message as a String.

\paragraph{Customization.}
\label{par:customizations}
The last field included in a \classname{HttpMessage} is the \textbf{customization} field, defined on line 15,  which in future will grow in complexity as it is the only way to integrate deep configurations into the code. It is essential, as it is used to satisfy the expectations regarding configuration that can be derived from user stories \ref{us:u3} and \ref{us:u5}. Since the urge for simplicity completely contradicts the aim of configuration (especially of minor details), the \classname{Customization} object is completely optional. One can either configure, or alternatively use the defaults which should be sufficient and work in the majority of scenarios.

The configurations included in the \classname{Customization} object have less to do with the message per se and more with the HTTP client in the background. One could argue to include countless different kinds customizations into the DSL. Here we give three customization examples to showcase how it works.
\begin{enumerate}
    \item The first customization option is to define a proxy server by using the \textbf{proxyServer} field in the \classname{Customization} object. The user enters a URL and a port, both of which can be set with a string or a variable. 
    \item The second customization option is to add basic access authentication, a standard form of authentication used for websites \cite{basicAuth}. For this the  \classname{BasicAuth} object is introduced. The user enters the credentials with a string or a variable, however since one does not want credentials in code that is potentially visible to the public, it is recommended to use environment variables for this purpose.
    \item The third and final option of customizations is to change the timeout. Here the user can set the maximum time interval (in milliseconds) that the client would wait for a response. In case no response comes within that timeframe, the client throws a timeout exception. The default value for this option is 5000, i.e., 5 seconds.
\end{enumerate}

\subsubsection{A Simple Usage Example}
\label{subsubsec:dslExa}
Given this DSL definition, the chosen example of use is a HTTP request towards a weather API provided by \textit{Accuweather}~\cite{accuwheather} in order to retrieve basic information about a location. The HTTP-DSL description for such a request can be found in \autoref{fig:dslEx}. In  \autoref{sec:application} this example will be discussed in detail, and one further example will be provided. 

\begin{itemize}
    \item 
To use the HTTP-DSL, a user has to first create a file with a "\textit{.http}" file extension, then open it and define HTTP messages by typing the keyword "http" and encapsulating the following commands with braces (the braces in lines 1 and 9 of \autoref{fig:dslEx}).

\item The user must then give the HTTP message a proper name. As shown in \autoref{fig:dslEx}, for this example the chose name is "WeatherLocation".

\item In line 3 we see the basic form of setting a URL: type the keyword "url" and  then set the two attributes \textit{server} and \textit{path}.\\
To set the server, write the keyword "server" followed by a valid URL. In the example, on line 3 we see that the URL is "\textit{http://www.dataservice.accuweather.com}". Also "\textit{www.dataservice.accuweather.com}" would lead to the same outcome, as HTTP is the default protocol for this DSL. The semantics behind this HTTP request are discussed in \autoref{subsec:weatherapp}, where this example will be extended. \\
The \textbf{path} field is set in line 4 by typing the keyword "path". The specified path for the example is "\textit{locations/v1/cities/search}".

\item 
The next field, \textbf{requestMethod}, is entered in line 5 by leading with the keyword "type" followed by GET. The alternative options POST, PUT and DELETE can be chosen from a list when pressing ctrl+space in an eclipse environment.

\item The next field used in this example is \textbf{query} in line 6 to 8. Here, three \textbf{parameters} are entered.
That is done by leading with the keyword "param" followed by the key-value pair, separated by a colon. The first key-value pair begins with the key "\textit{apikey}" followed by the colon and the value,  an \classname{Inputvariable}. The \classname{Inputvariable} is inserted by typing the "input" keyword followed by the variable name, here  \textit{apiKeyParam}. The value for this variable can be entered later in the process model. The second and third \textbf{parameter} are entered in a similar fashion, with the third one having a static input for the value, here "\textit{en-US}".
\end{itemize}

This example uses neither \textbf{headers} nor a \textbf{customization}, however, an implementation with all those possible options can be found in \autoref{app:dslExample}.

\subsection{Code Generation}
\label{subsec:codegenerator}

This code generator is used for all the Java source code that needs to be generated to run the HTTP clients which ultimately send the HTTP requests. However, the simple generation of a few Java classes is not enough, as that is not sufficient to fully integrate the code into a DIME application.\\  
In \autoref{subsubsec:httpClient} it is detailed how and what is generated for the HTTP clients. Then \autoref{subsubsec:responseHandler} explains how to generate the custom response handler, that is used for processing HTTP responses. The whole process of putting these generates together to finally integrate them into DIME is described in the last section, \autoref{subsubsec:mavenProject}.


\subsubsection{HTTP Client}
\label{subsubsec:httpClient}
The first component that needs to be generated is the HTTP client. For this purpose the Apache HTTP Client library running version 4.5.13 \cite{HTTPCLIENT} is used. The reasons for choosing this particular library are that this library is very well maintained by a big organization, and that most of the back-end code of DIME applications is written in Java or Xtend. Having additional code in java will be easier to maintain for DIME developers in the future.
The particular version 4.5.13 fits well, because it is Java 8 compatible and older versions of DIME can still use it.
The implementation of the code generator was done by using template based code generation with the help of Xtend templates. These are designed for code generation and make it effortless to optionally embed templates into other templates.

\begin{figure}[t]
	\begin{center}
		\includegraphics[width=\textwidth]{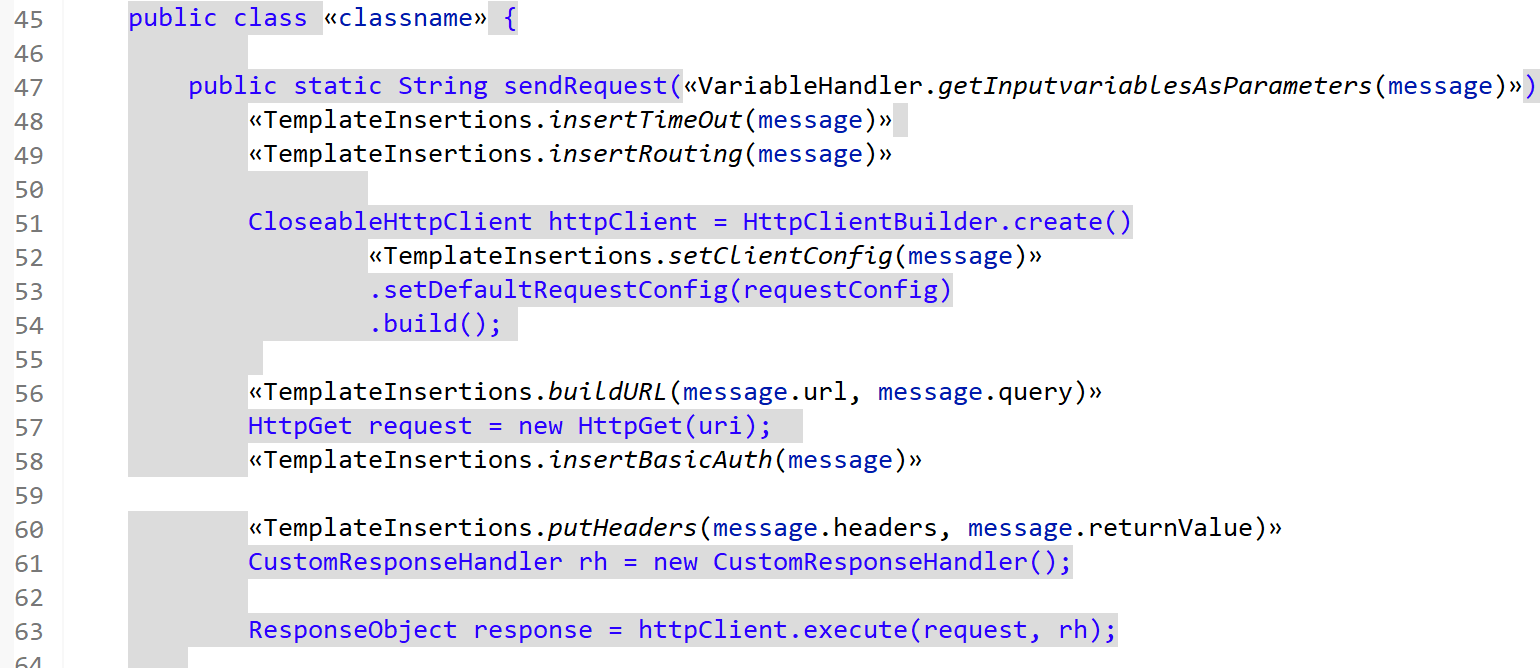}
		\caption{Snippet of the code generator template for HTTP clients doing GET requests.}
		\label{fig:get1}
	\end{center}
\end{figure}

In \autoref{fig:get1} one can see the first half of a template used for generating a class to send a HTTP message with a GET request method. \autoref{fig:get1} is a screenshot instead of a Listing in order to show that the Xtend templates provide a good overview: in the grey areas is the template code, that will be generated "as-is", and in the white areas we see the code insertions that call functions which insert code snippets that will be generated. 
\begin{itemize}
\item Line 47: one can see that the generated class will  have a single static method called sendRequest() that will be used to send the HTTP message and return the desired payload. The parameters of this method, if existent, are all derived from variables used in the DSL. Therefore, an external utility class called \classname{VariableHandler} has a method called \classname{getInputVariablesAsParameters} which takes the message that was previously configured in the DSL and creates a String consisting of all the variables used. This String is then taken and inserted into the template.
\item Lines 48-49: the second utility class called \classname{TemplateInsertions} is used to optionally insert template snippets for the timeout and proxy configurations if they are desired. This way the utility classes are used to check the message on each occasions for every option, and provide the template with code snippets when they are needed.
\item
Lines 51-54: the code for building the \classname{HttpClient} of the Apache library is generated. It is also dependent on the message, as can be seen in line 52.
\item 
Analogue further templates are inserted in lines 56, 58 and 60.
\end{itemize}

\subsubsection{ResponseHandler}
\label{subsubsec:responseHandler}
To control how the HTTP client reacts, a new class called \classname{Custom\-Response\-Handler} is also generated. For the full code see \autoref{app:CRH}. This class receives a \classname{org.\-apache.\-http.\-HttpResponse} object. The \classname{CustomResponseHandler} implements the \classname{org.\-apache.\-http.\-Response\-Handler} interface, thus it has the method \classname{handleResponse}. Within that method, the response is processed by returning a \classname{ResponseObject} that carries necessary information. The class \classname{ResponseObject}, of which an excerpt can be seen in \autoref{fig:responseObject}, is also generated and it is a plain old java object (POJO) holding the fields \textbf{payload}, \textbf{statuscode}, \textbf{succeeded}, \textbf{tryAgain}, \textbf{nextUri}, \textbf{requestType}. These fields are vital for the HTTP client, as those will indicate whether or not the \classname{sendRequest} method succeeded or whether it has to take further actions (by resending or redirecting the message, for example).\\

\begin{lstlisting}[
label={fig:responseObject},
style=java,
caption={\classname{ResponseObject} variable declaration.}
]
public class ResponseObject {
	private final String payload;
	private final int statuscode;
    private final boolean succeeded;
    private final boolean tryAgain;
    private final String nextUri;
    private final RequestType requestType;
\end{lstlisting}

\subsubsection{Maven Project}
\label{subsubsec:mavenProject}
The initial plan was to simply generate all the necessary classes into a preexisting DIME project and make them available for any service that might need it.

This approach was however not how things work within the DIME framework. If new plugins are inserted in DIME, they are standalone Maven Projects which will be Maven Modules within the big Maven Project that holds the entire DIME application.

\begin{figure}[t]
	\begin{center}
		\includegraphics[scale=0.9]{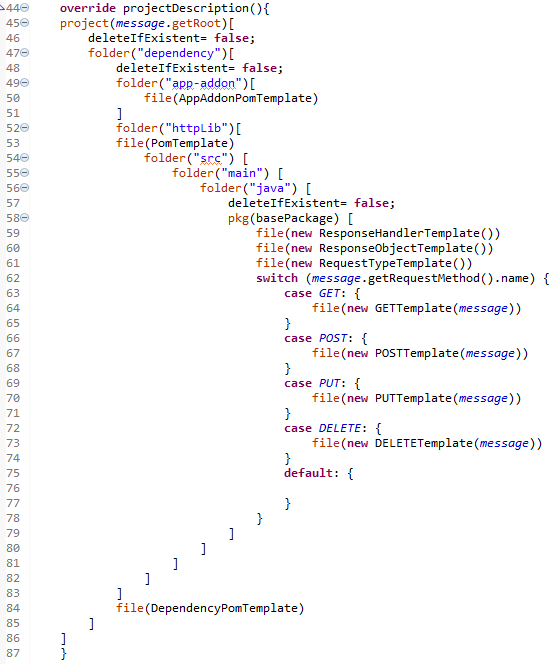}
		\caption{Project Structure of the generated Maven Project.}
		\label{fig:projectStructure}
	\end{center}
\end{figure}

To generate a Maven project, a DSL called \classname{project\-Description} \cite{projectDescription} is used. The \classname{project\-Description} defines the structure of project, meaning that one can easily design a folder hierarchy. Within these folders one can insert Xtext Templates, and this means that all the classes, previously mentioned in \autoref{subsubsec:httpClient} and \autoref{subsubsec:responseHandler} can be generated within the project structure.\\
The final structure of the generated Maven project can be seen in \autoref{fig:projectStructure}, which is a screenshot of the code since the design of the  \classname{projectDescription} is very easy to comprehend.

The Maven project for the HTTP client begins in line 52 and ends in line 83\footnote{The code in lines 45-52 and 84-86 is for integration purposes and will not be further discussed.}. The top level folder for the Maven project called \textit{httpLib} is set up in line 52.  Within that folder there are a file and a folder. The template for the file is defined in a class called \classname{PomTemplate} and it holds all the necessary templates to generate a standard \textbf{pom.xml} that can be found in Maven projects. The first folder within the \textit{httpLib} folder is called \textit{src} (line 54). The hierarchy goes down through the folders \textit{main} and \textit{java} where the actual code will lie. Within the \textit{java} folder a multitude of templates will be called: in line 59 the template for the \classname{CustomResponseHandler}, in line 60 for the \classname{ResponseObject} and in line 61 for an enum called \classname{RequestType} that is utilized in the \classname{ResponseObject}.
In lines 62-77 a switch-case command is used to call the correct template for the given HTTP message.

The \classname{projectDescription} DSL will be called for each single HTTP message used within the DIME application, however since in line 57 it is coded that existent code will not be deleted, every HTTP message template saved after the first one will simply be added to the existing file structure.
With this file structure the DIME framework will automatically take all the necessary measures to integrate the generated files into the application.

\subsection{Semantic Integration Into the DIME Context}
\label{subsec:semanticIntegratio}
Even though it was previously mentioned that the generated files are integrated into the DIME application, this is only partially correct. The generated classes will be readily available within the application during runtime, however, no component within the application is linked to these classes yet. To integrate the classes semantically, meaning that they will be meaningfully used, the  \textbf{generic SIBs plugin} previously introduced in \autoref{subsec: genericsibs} needs to be implemented.

\paragraph{IModelTrafoSupporter.}
The class \classname{HttpModelTrafoSupporter} is the implementation of \classname{IModelTrafoSupporter} and covers multiple objectives:
\begin{enumerate}
    \item It specifies that the \classname{generic SIB} plugin shall scan all the files with \textit{.http} file extensions for objects of type \classname{HttpMessage}. This way it  ensures that the user can ultimately define multiple HttpMessages per description file.
    \item An object of the class \classname{HttpSIBModellingProvider} is set as the \textbf{modellingProvider}.
    \item An object of  class \classname{HttpFileGenericSIBGenerator} is set as the \textbf{generationProvider}.
    \item The folder where the generated SIBs will be placed is named \textit{HttpSIBs}.
    \item Finally, a new icon is set.
\end{enumerate}

\newpage
\paragraph{HttpSIBModellingProvider.}
The next required class is \classname{HttpSIBModellingProvider}, which is an implementation of the \classname{IGenericSIBBuilder} interface. Within that class it is defined that:
\begin{enumerate}
    \item The name and label of the generated SIB are the \textbf{name} attribute of the \classname{HttpMessage} object.
    \item The input ports of generated SIB are all the \classname{Inputvariables} specified in the \classname{HttpMessage} object. This is done by using the same  \classname{VariableHandler} utility class  already used for the code generators. The method \classname{getInputVariables} is used to scan the message and return a list of all the \classname{Inputvariables} used. These are then mapped one by one to \classname{GenericPromitivePorts} and added to the list of input ports.
    \item The generated SIB will have two output branches called \textit{Success} and \textit{Failure}. The content of those branches is per default the payload of the response as a String.
\end{enumerate}

\paragraph{HttpFileGenericSIBGenerator.}
This final required class, called \classname{Http\-File\-Generic\-SIB\-Generator}, is the implementation of the \classname{IProcessSIBGenerator} interface. The contents of the methods were outsourced to another class\footnote{For simplicity, this will be ignored in this thesis as it has no impact on the functionality of the code}. The \classname{generateContent} method does two things: firstly it initiates the code generation of the HTTP clients (previously discussed in \autoref{subsubsec:httpClient}; secondly it generates the code for the \classname{sendMessage} method, that will call the HTTP clients during runtime. This method has rather confusing code in its template because the majority of code is outsourced, however the functionality can be explained quickly. With the \classname{HttpMessage} object and the \classname{VariableHandler} utility class (previously discussed in  \autoref{subsubsec:httpClient}), all the input variables are extracted and inserted as argument for the call of the \classname{sendRequest} Method that the HTTP clients provide. The value returned by the \classname{sendRequest} method is then taken and inserted into the branches \textit{Successful} and \textit{Failure}. \\
\classname{GetMethodCallSignature} is the method within the \classname{Http\-File\-Generic\-SIB\-Generator} that is returning the method signature of \classname{sendMessage}, which is the method that has just been generated. Without the \classname{GetMethodCallSignature}, the DIME application would not know how to call the method, meaning that none of the code that has been generated could be reached by the application.\\
Finally, the last method implemented is \classname{getResultTypeName}, which provides the type of response returned by the \classname{sendMessage} method.
\begin{figure}[h]
	\begin{center}
		\includegraphics[scale=0.9]{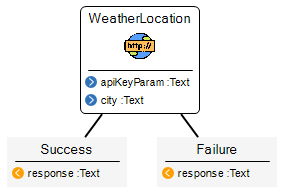}
		\caption{The SIB generated from \autoref{fig:dslEx}.}
		\label{fig:SIBEx}
	\end{center}
\end{figure}
\paragraph{Example SIB.}
\label{par:exampleSIB}
\textbf{Generic SIBs} take in a description file and generate a corresponding SIB. We use the description file from the previous usage example in \autoref{subsubsec:dslExa} to showcase how a generated SIB with the previously mentioned configurations looks like.  Given the description file, containing the code of  \autoref{fig:dslEx}, one can now generate the SIB shown in \autoref{fig:SIBEx} with the push of the save button.

In \autoref{fig:SIBEx} one can see that:
\begin{enumerate}
    \item the label of the SIB is the name attribute.
    \item the two \classname{Inputvariables} used in the file are now input ports of the SIB.
    \item the SIB has two output branches, \textit{Success} and \textit{Failure}.
    \item the outcoming response is of type \textit{Text}, which is the DIME representation of a String, because the default output value is the payload of the HTTP response as a String.
\end{enumerate}
If the user were to define a new HTTP request within the description file, by saving the file the user could automatically use a corresponding SIB of that request in a process model.

\subsection{Comparing With the Intellij IDEA HTTP Client}
Jetbrains \cite{jetbrains} is the company that provides one of the most popular IDEs: Intellij IDEA \cite{intellij}. Integrated into the IDE is a feature, the \textbf{HTTP Client plugin}, that can certainly be compared to the HTTP-DSL. This plugin uses a textual DSL that allows for quick definitions of HTTP requests that shall be used either to learn how to access an external web service or to test a web service that one has developed to check for errors. However, these HTTP requests can not be integrated in an application: they act as a standalone tool for the purposes mentioned above.

\paragraph{HTTP Client DSL.}
The language used to define HTTP requests within the plugin is in compliance with the definition and the characteristics of a DSL as previously discussed in \autoref{subsec:DSL}. Thus we refer to this language as the \textbf{HTTP Client DSL}.\\
The HTTP Client DSL allows for a very quick definition of HTTP requests. The first line of each HTTP request is used to state the name of the request. The next line begins with the request Type followed by the complete URL. After that, each following line can be used to define a header, resulting in overall very few lines of code needed. Similar to our HTTP-DSL, the HTTP Client DSL features environment variables and dynamic variables (the latter can be set via configuration files). \\
Custom response handlers can be set with script insertions and the HTTP Client DSL even allows for code injections with immediate support of code highlighting for these. 

\paragraph{Comparing the Code.}
When comparing the lines of code required for the definition of a HTTP request, the HTTP Client DSL comes out on top compared to the HTTP-DSL.\\
The example HTTP request in \autoref{fig:dslEx} was discussed in \autoref{subsubsec:dslExa}. To define the same request within Intellij it only takes the two lines of code reported in \autoref{fig:intellijDSL}. However, the few required lines of code come at a price. The second line is incredibly long, since the URL is not segmented into distinct parts, but simply written down as it would be in a web browser. This makes the code hard to read and harder to check for errors compared to the HTTP-DSL version.\\
Moreover, the HTTP Client DSL does not offer deep configurations, apart from the response handler.\\

\begin{lstlisting}[
label={fig:intellijDSL},
style=HTTP,
caption={HTTP Client DSL version of code seen in \autoref{fig:dslEx}.}
]
### WeatherLocation
GET http://dataservice.accuweather.com/locations/v1/cities/search?apikey={{$timestamp}}&q={{$city}}}&language=en-US
\end{lstlisting}

\paragraph{Concluding the Comparison.}
Even though both DSLs are clearly meant to describe HTTP requests, to a degree it is like comparing apples to oranges: their purposes differ to a large degree, thus the demands for each DSL are completely different.\\
The HTTP Client DSL is meant for testing, thus it will never be integrated into processes, rendering the option for configurations within that DSL completely useless. The HTTP-DSL, in contrast, is used to generate Java code that is to be deeply intertwined within the structure of a running web application. Sometimes these integrations into complex workflows might not be possible at all without a certain degree of configurability.\\
The HTTP Client DSL has one more feature that is great within its context but makes no sense at all to be included in the HTTP-DSL: The ability to open a given HTTP request in the web browser. While this feature is fantastic for testing certain features within a  web application, it would not be a useful addition to the HTTP-DSL: the HTTP-DSL will simply never ever call operations within an API that are designed to be opened on a web browser, since it is designed to make use of web services, and web services are by definition designed to be for machines only.\\

\newpage
\section{Application}
\label{sec:application}

The purpose of this section is to give an example of how the HTTP-DSL and its integration into DIME can be applied in real web applications.\\
When looking at the user stories in \autoref{subsec:Userstories} it is clear from \ref{us:u1}, \ref{us:u2} and \ref{us:u4} that most web developers want to use HTTP requests for simple API calls that sometimes require certain workflows. Consequently, it makes sense to have a simple DIME application that can showcase whether or not this is possible with the HTTP-DSL and whether the inclusion of the DSL in that process actually assists the developer.

\subsection{Using the HTTP-DSL Within DIME.}
\label{subsec:appStart}
This section introduces the steps needed to first create a DIME application with the HTTP-DSL plugin, then use the HTTP-DSL and integrate it into the application.

\begin{figure}[h]
	\begin{center}
		\includegraphics[scale=1]{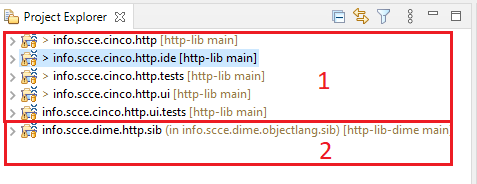}
		\caption{Files needed for HTTP-DSL plugin, found in the \textbf{Project Explorer}.}
		\label{fig:pluginFiles}
	\end{center}
\end{figure}

\paragraph{Setting up DIME.}
\label{par:setUpDIME}
To begin developing a DIME application, one first has to download and install the DIME package as described on their Gitlab wiki \cite{dimeInstall}. Usually the user would now simply start the \textit{dime.exe} and begin modelling the application. However, since the HTTP-DSL is not included in the core DIME package, the user has to download and import the necessary files for this plugin.
To import the files, right click in the \textbf{Project Explorer} (described in \ref{par:dimeLanguages}), click \textit{Import $\xrightarrow{}$ General $\xrightarrow{}$ Existing Projects into Workspace $\xrightarrow{}$ next}, and select the location of the downloaded files. When successful, it should look like in \autoref{fig:pluginFiles}. There, the folders in box 1  are used for the HTTP-DSL and its code generator. The folder in box 2 contains the files  needed to configure the Generic SIBs plugin.

\begin{figure}[h]
	\begin{center}
		\includegraphics[scale=0.6]{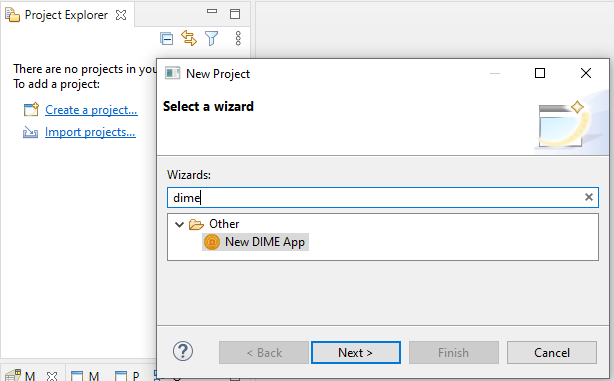}
		\caption{Creating a new DIME application. }
		\label{fig:installDIME}
	\end{center}
\end{figure}

The DIME user now has to right-click on one of the folders and click "\textit{Run As $\xrightarrow{}$ Eclipse Application}". Once the new Eclipse environment is ready, the user can use DIME normally but with the additional access to the new features.\\
To begin the design of a DIME application, the user has to click on \textit{Create a new Project..}, then search for "\textit{dime}" and select "\textit{New Dime App}", as in \autoref{fig:installDIME}. The user then has to select "create a new application from scratch" and proceed by clicking \textit{finish}.

\begin{figure}[h]
	\begin{center}
		\includegraphics[width=\textwidth]{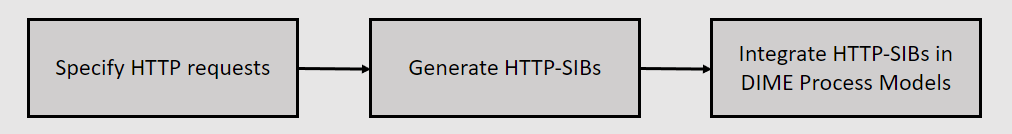}
		\caption{Three step process to create and integrate HTTP requests into DIME.}
		\label{fig:AppProc}
	\end{center}
\end{figure}

\paragraph{Using the HTTP-DSL.}
\label{par:useHTTPDSL}
When the DIME project is set up, the actual web application development can begin. To use the HTTP-DSL, the user has to go through the three step process  depicted in \autoref{fig:AppProc}.\\

Going from left to right, the developer first needs to:

\begin{enumerate}
    \item \textbf{Specify HTTP requests.} To specify the HTTP requests the developer first needs to create a new file (usually within the \textit{dime-models} folder) with a "\textit{.http}" file extension. The definition of a HTTP request begins with the keyword \textit{http} and a pair of braces. Within these braces, all the essential fields (see  \ref{subsubsec:essentialFields}) and optionally, some nonessential fields (see \ref{subsubsec:nonessentialFields}) have to be entered.
    \item \textbf{Generate HTTP-SIBs.} This step is quick but crucial. The description file needs to be saved, as that will trigger the generation of SIBs from the HTTP requests specified in the previous step.
    \item \textbf{Integrate HTTP-SIBs in DIME Process Models.} Now that the SIBs are created, the developer can simply drag and drop them into \textbf{ProcessModels} (see \ref{par:dimeLanguages}). If variables were used within the definition of a HTTP requests, the corresponding \textbf{input ports} of the SIBs need to be filled with data.
\end{enumerate}

\subsection{Application 1: The Weatherapp}
\label{subsec:weatherapp}
The DIME framework is built to make full stack development easy by coding the application in a model based approach. Therefore the final product will be a full stack web application with both front-end and back-end. As the name \textit{Weatherapp} suggests, the final product should provide the user with basic weather information for a desired location, extending the basic example introduced in \autoref{subsubsec:dslExa}. The location will be a city that the user specifies by entering a city name into the web application. The provided information will be basic weather data in form of a short sentence describing the current weather conditions for that city. This data is provided by the service \textit{Accuweather} via their API \cite{accuwheather} accessible by registered users with standard HTTP requests.\\
Since the focus of the thesis is not centered around the front-end, this application will have a very basic design that provides text inputs, text fields and buttons to show the functionality of the process that is taking place in the background. The main objective is to see the usage of the HTTP-DSL in the back-end built with DIME.

The implementation process follows the steps described in \autoref{par:useHTTPDSL}. Since the creation of an empty DIME application is not specific to this \textit{Weatherapp} example, we show in   \autoref{subsubsec:WeatherAppDescFile} how to carry out step 1 of the three steps process depicted in \autoref{fig:AppProc}: defining the HTTP requests in the HTTP-DSL description file. Since step 2 only consists of saving the file, we then show  how to integrate the HTTP-SIBs in the \textbf{ProcessModel} in \autoref{subsubsec:WeatherAppIntegration}. In  \autoref{subsubsec:WeatherAppFinal}, we show the final result of the web application. 

\subsubsection{The Description File}
\label{subsubsec:WeatherAppDescFile}

\paragraph{Weather API Functionality.}
\textit{Accuweather}'s API has a two step process for the use case mentioned above, apart from registering to obtain an API key.
\begin{enumerate}
    \item The developer calls the \textit{Locations API} \cite{accuwheatherLoc} to receive a location key that is unique to the specified location. Since the user will enter a city name, there will be conflict whenever multiple cities across the globe share a name. In that case, the application will simply pick the top city that matches the specified name, as this city will also have the highest population thus be the most likely requested.
    \item The developer then calls the \textit{Current Conditions API} \cite{accuwheatherCC}. This API takes the previously obtained location key and returns a javascript object in JSON format with information regarding the current weather conditions at the location specified by that key.
\end{enumerate}

\paragraph{Modeled API Requests.}
\label{par:WeatherAppDescFile}
Two HTTP requests will be needed, meaning two \classname{HttpMessage} objects inside the description file. To create a description file, the developer navigates to the \textit{dime-models} folder and creates a new file with \textit{.http} as file extension. 
For the WeatherApp, this description file includes the code shown in \autoref{fig:appDescFile}.

The first HTTP message is called \textit{WeatherLocation} and it is used to send the first request towards the API to retrieve the location key. Since the \textbf{server} and \textbf{path} fields for the URL are static, they are directly entered in lines 3-4 in the description file without using variables.\\[20mm]

\begin{lstlisting}[
label={fig:appDescFile},
style=HTTP,
caption={Weatherapp HTTP message description file.}
]
http{
	name "WeatherLocation"
	url server http://dataservice.accuweather.com
		path locations/v1/cities/search
	type GET
	param "apikey" : input $apiKeyParam
	param "q" : input $city
	param "language" : "en-US"
}
http{
	name "CurrentConditions"
	url server http://dataservice.accuweather.com
		path input $path
	type GET
	param "apikey" : input $apiKeyParam
}
\end{lstlisting}

\noindent Since the API will only work for authorized users, the developer has to transfer an API key via the \textit{apikey} parameter. The API key is static, however one should never leave credentials or API keys in version controlled code, thus a variable is used in line 6 to insert the key in the modeling phase.\\
The \textit{q} parameter in line 7 is used to emit the desired city. The concrete city is known to the application only at runtime, thus a variable with name \textit{$city$} is used.\\
The final component of the first message is the parameter \textit{language} in line 8. It should be English at all times, thus it is filled statically with the value \textit{en-US}.

The second HTTP message is called \textit{CurrentConditions}, starting in line 11. It is  implemented analogously, with one critical detail  different: the \textbf{path} field of the URL in line 13 is a variable. It is done this way because the path contains the location code of the city, which is only known at runtime. Thus it can not be entered statically in the description file.

\begin{figure}[!h]
	\begin{center}
		\includegraphics[scale=1]{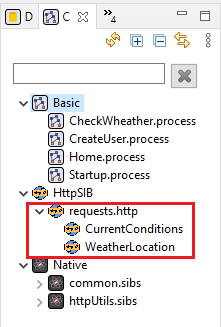}
		\caption{Palette of generated HTTP-SIBs found in the \textbf{ModelsView}. }
		\label{fig:WeatherAppSIBPalette}
	\end{center}
\end{figure}

Now that both  HTTP requests are defined, the file is saved to automatically generate the HTTP-SIBs into DIME. The generated SIBs palette is found in the \textbf{ModelsView} (defined in \ref{par:dimeLanguages}), as can be seen in \autoref{fig:WeatherAppSIBPalette} within the red square.

\subsubsection{Integrating the HTTP Requests Into DIME}
\label{subsubsec:WeatherAppIntegration}
\begin{figure}[!ht]
	\begin{center}
		\includegraphics[scale=0.85]{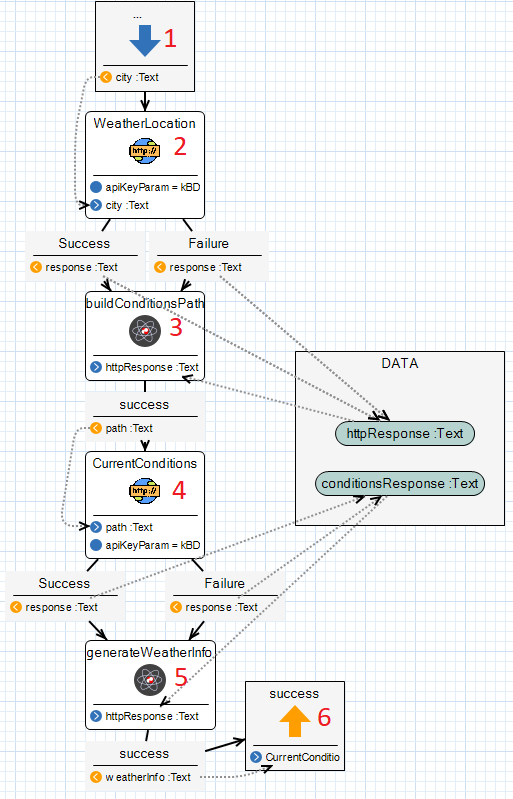}		\caption{ProcessModel of the checkWeather process. }
		\label{fig:WeatherAppPM}
	\end{center}
\end{figure}

The SIBs can now be draged and dropped into \textbf{ProcessModels} to create processes that define the business logic of the \textit{Weatherapp}. The whole process of taking a city name and returning the relating weather information is modeled in the \textit{checkWeather}  \textbf{ProcessModel}, shown in \autoref{fig:WeatherAppPM}. 

\paragraph{Introducing the ProcessModel Syntax.}
The syntax of \textbf{ProcessModels} is simple:
\begin{enumerate}
    \item The model begins with a single \textbf{start node} labeled with a blue arrow and ends with one or more \textbf{end nodes} labeled with an orange arrow.
    \item The control flow is modeled with directed \textbf{control flow edges}, depicted as solid arrows.
    \item The data flow is modeled with directed \textbf{data flow edges}, depicted as dotted arrows.
    \item \textbf{SIBs} are represented by rectangles with rounded edges, such as the nodes labeled 2, 3, 4 and 5 in \ref{fig:WeatherAppPM}. Each SIB has at least one outgoing \textbf{branch} that represents a possible outcome. A SIB can take in data via \textbf{input ports} (such as the one labeled \textit{city} in SIB 2) and return data via \textbf{output ports} (seen on the branches of SIB 2, below the branch names \textit{Success} and \textit{Failure}).
    \item The overall \textbf{data context} is represented by a square region labeled \textit{DATA} which contains variables, as seen on the right of \autoref{fig:WeatherAppPM}.
\end{enumerate}

\paragraph{The checkWeather ProcessModel.}
The control flow of the \textit{checkWeather} \textbf{ProcessModel}, shown in \autoref{fig:WeatherAppPM}, follows the nodes labelled in ascending order:
\begin{enumerate}
    \item The \textbf{start node} contains an \textbf{output port} that represents data entering  the \textit{checkWeather} SIB when it is used in a model models. The \textbf{output port} is labeled \textit{city} and is of type Text (a String). It will contain the city name that the application user will enter on the web page during runtime.
    \item The second node is the previously created HTTP-SIB  \textit{WeatherLocation}. It is used to retrieve the location key of a given city and has the two \textbf{input ports} \textit{apiKeyParam} and \textit{city} (defined earlier in \autoref{par:WeatherAppDescFile} when using variables). The API key is entered manually and the city \textbf{input port} is taken from the \textbf{output port} of the start node. The \textit{WeatherLocation} SIB has two possible outcomes and thus two branches: \textit{Success} and \textit{Failure}. To keep it simple, both branches are treated equally, saving the HTTP response in a variable called \textit{httpResponse} in the data context and continuing the control flow with the succeeding SIB. The HTTP response is a JSON object containing lots of details about given city, most importantly the \textit{location key} that is needed down the line.
    \item The SIB called \textit{buildConditionsPath} is a native SIB that can be created within DIME by referring to a Java method. This SIB takes the data from the \textit{httpResponse} variable, deserializes the JSON to retrieve the \textit{location key} and returns a String representing the path variable needed in the following SIB.
    \item The fourth node is the \textit{CurrentConditions} SIB that represents the second HTTP request. Its \textbf{input ports} hold the API key, that once again is entered manually, and the path that the previously discussed SIB returned. The \textit{currentConditions} SIB calls the API and returns the HTTP response containing a JSON object with lots of information regarding the weather conditions. This JSON is then saved in the \textit{conditionsResponse} variable, found in the data context, in form of a String. Once again both branches are treated equally, resulting in the next node.
    \item The \textit{generateWeather} SIB is once again a native SIB. It converts the JSON stored in the \textit{conditionsResponse} variable and returns a String with a short two to three words description of the weather. This description is then sent to the next node.
    \item The final node is per definition an \textbf{end node} and it has an \textbf{input port}, meaning that the \textit{checkWeather} SIB returns data. The data in this case is the previously created weather description.
\end{enumerate}

\subsubsection{Complete Application}
\label{subsubsec:WeatherAppFinal}
\begin{figure}[h]
	\begin{center}
		\includegraphics[scale=0.75]{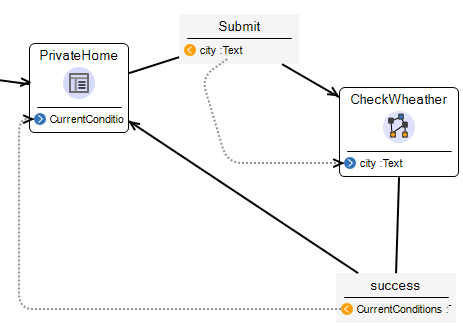}
		\caption{Overall logic of the \textit{Weatherapp}. }
		\label{fig:WeatherAppLogic}
	\end{center}
\end{figure}

Now that the \textit{checkWeather} SIB is finished, it is integrated in the overall business logic of the app, shown in \autoref{fig:WeatherAppLogic}.

The  \textit{PrivateHome} on the left is a GUI-SIB containing the front-end components of the application. The logic within that SIB will not be further discussed in detail, however for this example it is only relevant that it contains a text input field where to enter a city name and a submit button.
The \textit{PrivateHome} SIB has a \textbf{branch} that is triggered when the submit button is pressed, sending the city name to the previously discussed \textit{checkWeather} SIB. This SIB then returns the short weather info text through the \textbf{output port} called \textit{currentConditions} back to the \textit{PrivateHome} SIB to diplay it.

\begin{figure}[h]
	\begin{center}
		\includegraphics[scale=0.9]{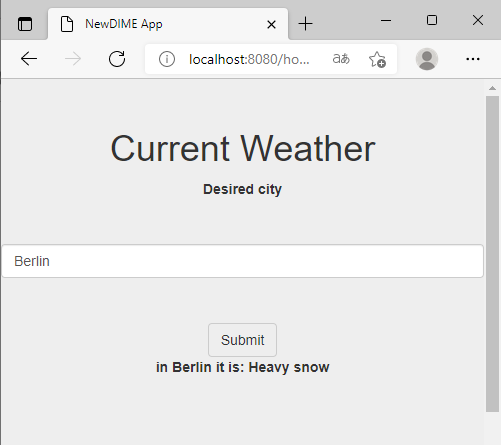}
		\caption{Appearance of the \textit{Weatherapp} in a browser. }
		\label{fig:WeatherAppWebBrowser}
	\end{center}
\end{figure}

\paragraph{Running the Application.}
The final application, when run in a web browser, looks as depicted in \autoref{fig:WeatherAppWebBrowser}. The user interface is fairly basic, offering a text input and a submit button. The weather information, if form of a small sentence, is printed below the submit button.

\subsection{Application 2: Defining a REST Layer}
\label{subsec:RESTLayer}
The HTTP-DSL provides a very basic communication layer that can be used to abstract even further. How to use  the further abstraction approach will be demonstrated on the example of the Representational State Transfer (REST) \cite{Fielding2000} layer. The REST architectural design was created to offer a set of simple and standardized communication primitives for services within the World Wide Web to increase their ease of communication and their reliability. The term \textit{RESTful API} is commonly used to describe APIs that are stateless, meaning that any response from the API can be interpreted without (or independently of) a context.

\paragraph{Implementation of the REST Layer.}
\label{par:RESTImpl}
The implementation of the REST layer is a basic proof of concept useful to showcase how additional abstractions from the HTTP-DSL can further increase productivity. The main idea of this REST layer is to provide a set of template requests that do not have to be defined each time within a HTTP-DSL description file, but already exist as ready to use SIBs within the \textit{ModelsView} (see \ref{par:dimeLanguages}) of DIME.
For that purpose, the implementation of the REST layer only includes the four HTTP requests representing the basic GET, POST, PUT and DELETE request types used to access a REST API. They are defined in a description file that makes use of the HTTP-DSL primitives. The four are implemented in almost identical fashion, thus we discuss here only the GET request definition, seen in \autoref{fig:RESTGetDescFile}, as a representative. The full code for the four requests can be found in \autoref{appendix:RestDSLFile}. \\[5mm]

\begin{lstlisting}[
label={fig:RESTGetDescFile},
style=HTTP,
caption={DSL code for the GET request.}
]
http{
    name "GET Request"
    url server input $url path input $path
    type GET
    param input $param1Key : input $param1Value
    param input $param2Key : input $param2Value
    header input $header1Name : input $header1Value
    header input $header2Name : input $header2Value
}
\end{lstlisting}

\begin{figure}[h]
	\begin{center}
		\includegraphics[scale=1]{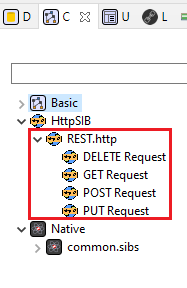}
		\caption{REST SIB palette in the DIME \textbf{ModelsView}.}
		\label{fig:restPalette}
	\end{center}
\end{figure}

In the GET request code reported in \autoref{fig:RESTGetDescFile} we can see that most of the fields, such as \textit{url}, \textit{headers} and \textit{params} are kept unspecified by simply using variables. This allows the user to put these details in the SIB input ports when modeling the specific applicationwithin the \textbf{ProcessModel}.
The requests, when saved, are transformed into the REST SIB palette shown in \autoref{fig:restPalette} in the DIME \textit{ModelsView}, that is available to DIME developers.

\begin{figure}[t!]
	\begin{center}
	~\\[-150mm]
		\includegraphics[scale=0.9]{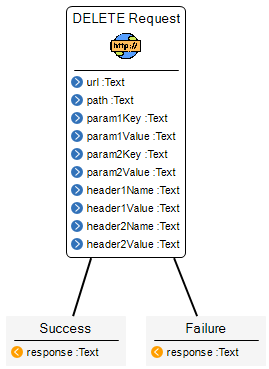}
		\caption{GET template request SIB within a \textit{ProcessModel}.}
		\label{fig:restWithinProcess}
	\end{center}
\end{figure}

DIME users can now comfortably drag and drop the desired request SIB into their process, and fill in the template with the necessary information. \autoref{fig:restWithinProcess} shows the {\tt GET request} SIB in DIME, which is such a template for a simple GET request. The SIBs for the POST, PUT and DELETE requests look almost identical, with the POST and PUT SIBs having an additional \textbf{input port} to insert a payload for the body.

One can see \textit{input ports} in the \textit{GET Request} SIB for almost all the necessary information. The user can now fill out all the ports needed either statically, by typing them in, or by assigning them data from the \textit{ProcessModel}.

\clearpage
\newpage
\newpage 
\section{Evaluation}
\label{sec:evaluation}

In this section, we discuss and evaluate the functionality, usefulness and overall success of the HTTP-DSL implementation. We start in \autoref{subsec:efficiency} with its efficiency, and how it could or should be applied in software development projects. In the \autoref{subsec:leanSoftwareDevelopment}, we then discuss the implementation and design of the HTTP-DSL from the perspective of Lean Software Development, as introduced in 2003 by Mary and Tom Poppendieck~\cite{PoppendieckPoppendieck03}. Finally, in  \autoref{subsec:finalConclusion} we evaluate the HTTP-DSL by comparing the user stories from \autoref{sec:Requirements}, that play the role of abstract requirements, with what the actual implementation provides.

\subsection{Evaluating the Efficiency of the HTTP-DSL}
\label{subsec:efficiency}
One of the main advantages that the HTTP-DSL brings to DIME is an increased velocity of development regarding HTTP requests. To prove this and to understand how the HTTP-DSL could and should be used in the long run, we analyze here two aspects: a basic comparison to a standard Java implementation of a HTTP request in \autoref{subsec:ComparingEfficiency}, then in  \autoref{subsec:furtherEfficiency} a discussion about how to further increase the efficiency of working with the HTTP-DSL.

\subsubsection{Comparison With a Traditional Implementation}
\label{subsec:ComparingEfficiency}
Here we evaluate the difference in productivity, measured in lines of code, when starting from scratch with the HTTP-DSL vs. a standard Java code solution. This standard solution uses the Apache HTTP Client \ref{subsubsec:httpClient} to implement  HTTP-requests.\\
The task for both implementations is a minimal GET request to a local API, to retrieve a list of users. This example is representative because this kind of requests is used fairly frequently in the context of web applications.\\

\begin{lstlisting}[
label={fig:quickSearch},
style=HTTP,
caption={Example request to a local API using the HTTP-DSL.}
]
http{
	name "AccessUserApi"
	url server localhost:8080 path users/all
	type GET
}
\end{lstlisting}

\paragraph{The Comparison.}
The shortest (still readable) format to implement this simple GET request with the Apache HTTP Client takes 30 lines of Java code, as shown in \autoref{fig:slowSearch}. 
When implementing the same request (with arguably more features in the background) within the HTTP-DSL, just the 5 lines of code of \autoref{fig:quickSearch} suffice.\\ 
With each added feature to the request the absolute difference of lines of code between source code and HTTP-DSL code can only increase.\\ 
The sizeable difference in lines of code is partly due to the amount of boiler plate code used in the traditional implementation. Lines 1-11  in \autoref{fig:slowSearch} do not include any semantics regarding the HTTP request, apart from the return type \textit{String} in the signature (line 11).\\

\begin{lstlisting}[
label={fig:slowSearch},
style=java,
caption={Example request to a local API using standard source code.}
]
package testApp;
import org.apache.http.HttpEntity;
import org.apache.http.HttpHeaders;
import org.apache.http.client.methods.CloseableHttpResponse;
import org.apache.http.client.methods.HttpGet;
import org.apache.http.impl.client.CloseableHttpClient;
import org.apache.http.impl.client.HttpClients;
import org.apache.http.util.EntityUtils;
import java.io.IOException;

public class HttpClientExample1_1 {
    public static void main(String[] args) throws IOException {
        CloseableHttpClient httpClient = HttpClients.createDefault();
        try {
            HttpGet request = new HttpGet("https://localhost:8080/users/all");
            CloseableHttpResponse response = httpClient.execute(request);
            try {
                HttpEntity entity = response.getEntity();
                if (entity != null) {
                    String result = EntityUtils.toString(entity);
                    System.out.println(result);
                }
            } finally {
                response.close();
            }
        } finally {
            httpClient.close();
        }
    }
}
\end{lstlisting}

Additionally, especially when looking at customizations (\autoref{par:customizations}) within the HTTP-DSL, simple features that in the DSL are set up with a single line of code take instead multiple alterations to the actual Java code. The benefits of the higher abstraction level provided by the HTTP-DSL are a quicker and easier implementation.

However the higher abstraction certainly has the drawback of inferior capabilities when fine tuning the code.

\subsubsection{Further Increasing Development Efficiency}
\label{subsec:furtherEfficiency}
The comparison in the previous section considered the code size when starting from scratch. However, the HTTP-DSL could be used differently, especially in the long run.\\
The REST layer implementation, seen in \autoref{subsec:RESTLayer}, demonstrates how further implementation time can be saved when stacking such an abstraction layer on top of the HTTP-DSL. The functionalities of the REST requests in the example are very basic, however, it could still have a significant impact. By providing a palette of REST requests, developers will no longer need to write any code for basic HTTP requests. These kinds of basic HTTP requests are used fairly frequently in web applications. An example for that would be the \textit{Accuweather} ~\cite{accuwheather} API. The whole \textit{checkWeather} process that is depicted in \autoref{fig:WeatherAppPM} could have been modeled without a custom HTTP-DSL description file. The \textit{GET Request} (introduced in \autoref{par:RESTImpl}) would be sufficient to model the two requests \textit{WeatherLocation} and \textit{CurrentConditions}, discussed in \autoref{subsubsec:WeatherAppIntegration}. This would allow for a solution that apart from the two native SIBs \textit{buildConditionsPath} and \textit{generateWeatherInfo} would not require any additional hand written code, further increasing efficiency within DIME. 

\subsection{Evaluating the Software Development Quality}
\label{subsec:leanSoftwareDevelopment}
The book \textbf{Lean Software Development: An Agile Toolkit} \cite{PoppendieckPoppendieck03} deals with the application of the \textbf{lean principles} in software development. When the book appeared, these lean principles were not  entirely new to the software development community, but the collection of these principles gave readers a tool kit that can be applied at various points during the software development process.

\subsubsection{The Lean Principles}
\label{subsubsec:LeanPrinciples}
The seven \textbf{Lean Principles} introduced and discussed in the book should act as guiding ideas in many disciplines. These principles, in contrast to concrete practices, are not meant as specific guides in certain situations. They are higher-level attitudes, that can help readers or practitioners to get into the right mindset, enabling them to recognize and resolve issues in the process of developing software. The seven \textbf{Lean Principles} are:
\begin{enumerate}
    \item \textbf{Eliminate waste:} Waste are all the things that do not increase the value of the product for the customer. That could even include unused features or changing the development team. As can be read in the book \cite{PoppendieckPoppendieck03} on page 13, "Whatever gets in the way of rapidly satisfying a customer's needs is waste". 
    \item \textbf{Amplify Learning:} In contrast to production, development is always about discovery. Hence, there is no streamlined process that one can apply for every software development project. Every project is individual, thus it is vital to amplify learning to get to the final product as quickly as possible.
    \item \textbf{Decide as late as possible:} Especially in domains with uncertainty, it is important to be agile. One tool for achieving this is to keep a structure that allows for late decision making. If done correctly one ends up with a complex system with the capacity for change. This way decisions can be made when the future is closer and easier to predict.
    \item \textbf{Deliver as fast as possible:} The typical development cycle goes: Design, implement, feedback, improve. The sooner one can deliver the product, the sooner one can also improve the product and the more can be learned. 
    \item \textbf{Empower the team:} Involving the developers with the details of technical decisions can improve the outcome tremendously as they are the ones most in touch with the minute details. Moreover, good leadership enables front line developers to make the best technical decisions. Good tools for that are pull techniques (e.g. daily meetings, comprehensive testing or visible charts) that allow for signaling between workers so that they can let each other know what needs to be done. 
    \item \textbf{Build integrity in:} Software with integrity is coherent allowing for easy maintenance, adaptions and extensions over time. This allows for the software to grow to a level where customers think that it suits their needs perfectly.
    \item \textbf{See the whole:} Maximizing the efficiency or productivity of one area in the software structure will often lead to a decline in another. It is vital to keep balance by allocating resources efficiently throughout the whole software project. This is a typical leadership problem as front line developers will typically first attend to their specialized interests.
\end{enumerate}

\subsubsection{Assessing the HTTP-DSL With Respect to the Lean Principles}
As DIME is a framework used for software development, more precisely web development, it is interesting to consider whether the HTTP-DSL extension of DIME improves DIME in regards to the \textbf{Lean Principles} summarized in \autoref{subsubsec:LeanPrinciples}.

\paragraph{Eliminating Waste.}
One of the features of the HTTP-DSL is the absence of all the boiler plate code typically used to implement a HTTP client in programming languages like Java. This fits perfectly into the criteria of \textbf{eliminating waste} as boiler plate code takes time and effort without any impact on the customer. Furthermore, it is very well possible to create not only process SIBs, but also GUI SIBs (see \ref{par:dimeLanguages}) from the HTTP-DSL description files. This way developers would not need to learn how to implement HTTP clients for multiple languages, but would only need to learn one language: the HTTP-DSL. Relearning new implementations would be waste, in the sense that it takes developer time without value increase for the customer.

\paragraph{Amplify Learning/ Empower the Team/ See the Whole.} 
DIME as a whole and the HTTP-DSL within it are very well suited for the \textbf{amplification of learning} in the development process. This is due to the fact that DIME users program by modeling. Modeling allows for an easier understanding of logic, connections and relations within the code as it is quicker and easier to interpret models than source code.\\
This has the additional feature that it is easier to \textbf{see the whole}, since exactly the connections between components within a project are generally hard to keep track of. The HTTP-DSL is integrated into DIME via SIBs that can then be inserted and arranged in models to create complex workflows. This, again, makes it easy to understanding the program and could potentially \textbf{empower the team} because other developers make decision with full awareness of any of the easy to comprehend HTTP request related code.

\paragraph{Decide as late as possible.}
Furthermore, the HTTP-DSL is designed to allow variable inputs which can be inserted when modeling or during runtime. Regarding the principle of \textbf{deciding as late as possible}, this feature checks all the marks. A perfect example is the decision on a host name or IP address within a HTTP client. Often developers do not know the IP addresses (or host names) of web services (e.g. within a local network) until they are set up to start the application. The HTTP-DSL features the use of environment variables, allowing the developer to specify details like the IP address on startup.
 
\paragraph{Deliver as fast as possible.}
Web applications often make use of external web services. The connection to these web services is usually implemented via the HTTP, resulting in the use of many HTTP request for each web application.\\
As discussed in \autoref{subsec:ComparingEfficiency}, when starting from scratch, the HTTP-DSL tremendously decreases the lines of code for standard HTTP requests. Even further gains in efficiency can be achieved, as seen in \autoref{subsec:furtherEfficiency}, meaning that the time spent to implement and use HTTP requests can be greatly reduced. This allows for a \textbf{faster delivery} of the overall program.

\subsubsection{Conclusion}
Overall it is safe to say that the HTTP-DSL can increase the productivity of DIME developers. It allows for a better understanding of the program, quicker implementation of HTTP requests and a decrease in "waste" along the lines of Poppendieck \cite{PoppendieckPoppendieck03} page 13. The HTTP-DSL can be regarded as a success with respect to the \textbf{Lean Principles}, as it addresses six of the seven principles described in the book. 
Hence, the HTTP-DSL contributes to the \textbf{Lean Software Development} nature of DIME and helps increasing the productivity of DIME users.


\subsection{Evaluation with Respect to the User Stories}
\label{subsec:finalConclusion}
This subsection evaluates the overall success of the HTTP-DSL as a solution with regards to the requirements raised in \autoref{sec:Requirements}. To this aim we will revert to \autoref{tab:grouping} and evaluate which demands of the user stories are met in each category. The outcome is summarized in \autoref{tab:eval}, where a tick expresses that the requirement is met.

\begin{table}[h]
	\begin{center}
 		\begin{tabular}{|c || c | c | c | c | c |}
			\toprule
			User Stories & Simplicity & Complex & Configuration & Reusability& Single Source  \\ 
			& & Workflows 
			& & & of Truth \\
 			\midrule
 			\midrule
 			\ref{us:u1} & X (\checkmark)&               &               &  &      \\
 			\ref{us:u2} &               & X (\checkmark)&               &  &      \\
  			\ref{us:u3} & X (\checkmark)& X             & X             &  &     \\
   			\ref{us:u4} & X (\checkmark)&               &               &  &      \\
 			\ref{us:u5} &               & X (\checkmark)& X (\checkmark)&  &     \\
 			\ref{us:u6} &               &               &               &  &  X (\checkmark)  \\
 			\ref{us:u7} & X (\checkmark)&               &               &  &   \\
 			\ref{us:u8} &               &               &               &  &  X (\checkmark)   \\
 			\ref{us:u9} &               &               &               &  &  X (\checkmark)  \\
 			\ref{us:u10}& X (\checkmark)&               &               & X (\checkmark)  &        \\
 			\ref{us:u11}&               &               &               & X (\checkmark)& X (\checkmark)\\
 			\ref{us:u12}&               &               &               & X  &      \\
 			\ref{us:u13}&               &               &               & X & X (\checkmark)     \\
 			\ref{us:u14}& X (\checkmark)&               & X (\checkmark)&   &    \\
			\bottomrule
		\end{tabular}
		\caption{Evaluation of success implementation with regards to \autoref{tab:grouping} .}
	\end{center}
\end{table}
\label{tab:eval}

\newpage
\paragraph{Evaluating Simplicity.}
\label{par:EvalSimplicity}
\textbf{Simplicity} is arguably the core requirement for this project, as it covers six unique user stories. The user stories \ref{us:u1}, \ref{us:u3} and \ref{us:u4} demand basic HTTP requests where no further configurations need to be made. The HTTP-DSL offers easy access to default solutions, either by defining a request in as little as five lines of code (see Sect.  \ref{subsec:ComparingEfficiency}) or by having default HTTP SIBs (see Sect. \ref{subsec:furtherEfficiency}) that do not require coding at all.
The simple and domain focused nature of the DSL should satisfy \ref{us:u7}'s demand for \textit{semantics over syntax} as well. Additionally, any HTTP request that has been defined in a description file can be used multiple times as a SIB in \textbf{ProcessModels}, satisfying the reusability demanded by \ref{us:u10}.
The final user story in the simplicity category is \ref{us:u14}. It is addressed by introducing environment variables, as exposed in \autoref{subsubsec:variables}.

\paragraph{Evaluating Complex Workflows}
\label{par:EvalComplexWorkflow}
The capability of integrating the HTTP-DSL into a \textbf{complex workflow} can be seen in \autoref{subsec:weatherapp}. Even though the \textit{Weatherapp} is not a complex application, one can see that HTTP requests can easily be used and combined in \textbf{ProcessModels}.

If a process demands for more advanced configurations of a HTTP request, as hinted by \ref{us:u5}, the developer can further customize it, as seen in \autoref{par:customizations}. These configurations are limited compared to code level solutions, however, the HTTP-DSL can and will be upgraded in the future to facilitate more and more needs in this regard.\\
One could reformulate the user story \ref{us:u3}  to "\textit{As an app developer I do not want to handle status codes, unless I need to do it}". Currently, the developer cannot change how the HTTP client reacts to certain status codes. Thus, this user story is labeled as successful with regard to the "\textit{I do not want to handle status codes}" version. However, when considering complex workflows, the handling of status codes is in the focus. This is why this demand is not yet met. However, it is possible to introduce feature of handling status codes into the HTTP-DSL in the future.

\paragraph{Evaluating Configuration.}
\label{par:EvalConfig}
As previously discussed in paragraph, developers can \textbf{configure} the HTTP requests.  One easy  addition to that is the seamless integration of HTTPS: it is as easy as leading the URL with "\textit{https}". Furthermore, it was discussed that user story \ref{us:u14} was directly addressed by introducing environment variables in the DSL. Thus, these two requirements are met. However, the third requirement is not yet implemented:  the requirement of handling status codes demanded by \ref{us:u3} is not yet possible, as was mentioned in the previous paragraph.
\paragraph{Evaluating Reusability.}
\label{par:EvalReusability}
The aspect of \textbf{reusability} has major importance, since it reduces the inefficiency of doing the same task over and over again, as discussed in \autoref{subsubsec:LeanPrinciples}. Thus, the user stories \ref{us:u10} and \ref{us:u11} are immediately addressed. As discussed previously in \autoref{subsec:efficiency}, two kinds of reusability are possible with this implementation. First, HTTP requests that are defined in the HTTP-DSL could be reused as SIBs in \textbf{ProcessModels} within the same projects. Alternatively, reusal can be taken a step further by offering packages of HTTP requests that can be used in any future DIME application.\\
The two demands (from \ref{us:u12} and \ref{us:u13}) in the \textbf{reusability} category that are right not yet met, would be possible to implement, if more time were available. Introducing the HTTP-DSL natively into DIME, as demanded by \ref{us:u12}, will come down to checking and testing whether the implementation is free of errors. The HTTP-DSL is currently used to generate backend code, however, with an additional \textbf{code generator} (like the one described in \ref{subsec:codegenerator}) one could also produce \textbf{GUI SIBs}, defined in Section \ref{par:dimeLanguages}, that could be used in the frontend. 

\paragraph{Evaluating Single Source of Truth.}
\label{par:EvalSSOT}
DIME and the HTTP-DSL within it are following the XMDD \ref{par:xmdd} paradigm. Every bit of code for the final web application is generated, meaning that if the code is not working, one can find the problem in the models (assuming that the framework itself is working correctly). Thus, the demands of user stories \ref{us:u6}, \ref{us:u8} and \ref{us:u9} are perfectly met.\\
The requirement of \ref{us:u13} to be able to generate frontend and backend code of the HTTP-DSL is technically not possible, as previously discussed. However, since the HTTP-DSL offers the capabilities in the realm of SSOT, as the DSL could work as a template for frontend SIBs as well.\\
The final demand of user story \ref{us:u11} is also met. The architecture of this solution, shown and discussed in  \autoref{sec:Implementation}, allows for an easy exchange of underlying code generators to adapt them in terms of used programming languages and libraries used.

\newpage
\section{Future Prospects}
\label{sec:futureProspects}
In this section the HTTP-DSL will be put into a bigger picture, allowing for an evaluation of possible enhancements or use cases for it. In \autoref{subsec:creatingSIBTemplates} the idea of providing ready-to-use HTTP-SIBs will be expanded, revealing the significance of that mechanism in the long run. In the second section, \autoref{subsec:integratingDataTypes}, the idea of actively integrating data objects into the HTTP-DSL is discussed, showing how it could be done and how much of an impact it would have on DIME.

\subsection{Creating SIB Templates}
\label{subsec:creatingSIBTemplates}
The idea of creating a palette of HTTP-SIBs was introduced in \autoref{subsec:furtherEfficiency} when discussing how the REST SIB palette could increase the efficiency when using HTTP requests within DIME. This idea of providing a set of HTTP Templates in form of SIBs can be taken even further.

\subsubsection{Increasing Granularity}
\label{subsubsec:increasingGranularity}
The practice of stacking more abstract layers on top of the HTTP-DSL could be regarded as a coarsening of its granularity. The REST layer is the most basic layer imaginable, as it only takes basic GET, POST, PUT and DELETE requests and provides them as SIBs within DIME. However, it would also be possible to not only provide single requests, but even whole  ready-to-use processes in form of SIBs for all future DIME users. 

\paragraph{Providing Complex Workflows.}
\label{par:OAuth2}
One example for that is OAuth2 \cite{OAuth2}, an industry-standard protocol for authorization. For OAuth2 to work, developers usually need to take certain steps that involve six requests and responses between the application user and multiple servers. Implementing this workflow is not trivial, in fact it involves quite a bit of research followed by a complex implementation with a multitude of aspects that can go wrong.\\ However, the workflow  needed for OAuth2 is always the same. What differs are a few parameters such as multiple URLs, an ID and a few more details. The whole process of OAuth2 could be modeled in DIME just once by a DIME developer, creating a SIB that has the full OAuth2 functionality with a set of \textbf{input ports} that represent the required parameters. 

\paragraph{A HTTP library.}
\label{par:HTTPLIB}
Implementing web applications within DIME would certainly be more efficient if the core DIME framework would be shipped with a library, similar to the \textit{common SIBs} \footnote{\textit{Common SIBs} are a palette of basic application domain-independent operations such as "TextEquals", which checks two Strings for equality, that are provided directly with DIME.}, containing a set of ready-to-use solutions including processes like the OAuth2 example mentioned above in \autoref{par:OAuth2}. 
Future DIME users would no longer have to come up with the ideas regarding such solutions ever again, leaving them more room for focusing on the unique aspects of their applications.

\subsubsection{Automating API Access}
\label{subsubsec:AutoAPIAccess}
The ideas presented above in \autoref{subsubsec:increasingGranularity} are very well suited for workflows that are always the same: in such cases, one template with a set of parameters covers all possible solutions. One of the aspects that cannot be regarded as "\textit{always the same}" within web applications and web services are APIs. Almost all the available web APIs are unique, since they carry unique data and offer unique services. This, however, does not mean that every request towards APIs has to be hand made.

\begin{figure}[!t]
	\begin{center}
		\includegraphics[scale=1]{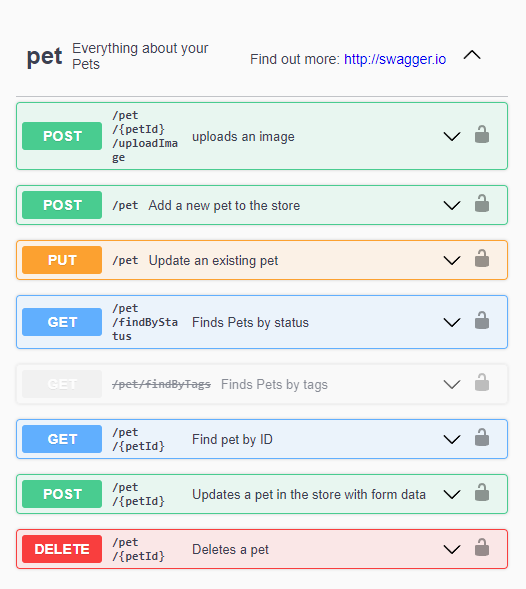}
		\caption{PET Store example: the SwaggerUI API documentation tool.}
		\label{fig:SwaggerUI}
	\end{center}
\end{figure}

\paragraph{Introducing OpenAPI.}
The \textbf{OpenAPI Specification} offers a standard for \textbf{RESTful APIs} that is designed to ease the understanding and creation of APIs without the demand for source code or documentation. The OpenAPI Initiative is a vendor-neutral project~\cite{OpenAPI2} that originated in the \textbf{Swagger} \cite{swagger} project owned by \textbf{SmartBear} \cite{smartbear}. As a project, Swagger is deeply intertwined with OpenAPI and offers a wide range of tools for designing, documenting, developing, and testing HTTP web services. One of them,  SwaggerUI, offers automatic documentation for APIs that follow the OpenAPI standard.

\paragraph{SwaggerUI Pet Store Example.}
\label{par:petstore}
The SwaggerUI documentation is depicted in \autoref{fig:SwaggerUI}, which is an excerpt from one of the official examples~\cite{Swaggerui} provided by Swagger. This SwaggerUI documentation displays in a browser the API of a potential pet store. It shows all the GET, POST, PUT and DELETE operations in unique colors by type, providing a quick overview. The tabs of these operations can be double clicked to get further information that includes parameters, response objects and the ability to test the API in the web browser.

\begin{figure}[!ht]
	\begin{center}
		\includegraphics[scale=1]{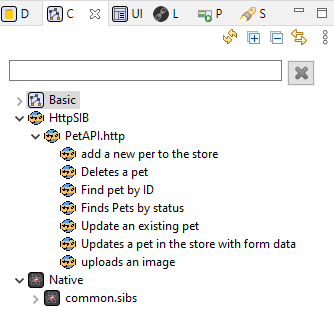}
		\caption{SIB palette matching PET Store API.}
		\label{fig:PETSIBS}
	\end{center}
\end{figure}

\paragraph{Using OpenAPI Documentation for DIME.}
Swagger offers tools that can take OpenAPI-conform APIs and create a documentation similar to SwaggerUI but machine readable. This documentation could be used to automatically generate a SIB palette, similar to the ones discussed in \autoref{par:HTTPLIB}, that is tailored to a specific API. \\
For the pet store example, the generated library could take the shape of the example depicted in \autoref{fig:PETSIBS}. Each  operation of the API is matched by a SIB named after the description of the operation. The \textbf{input ports} of these SIBs would be automatically mapped to the required parameters of the operations, so that  the users would only need to drag and drop the SIB in their process, link the required data and be done.

\newpage

\subsection{Integrating Data Types}
\label{subsec:integratingDataTypes}
One of the main drawbacks of the current version of the HTTP-DSL is its return value. Right now, the response is either a String containing the payload of the response, or the complete \classname{HttpResponse} object. This leads to the clear disadvantage that almost all the \textbf{HTTP-SIBs} must be followed by a \textbf{native SIB} that processes the data to make it viable. 

\subsubsection{Generating Domain Objects}
As described in \autoref{subsec:dime}, DIME developers can define within the \textbf{Data Language} \textbf{domain objects} that represent the data layer within the application. In the current version of the HTTP-DSL, it is not possible to send nor receive \textbf{domain objects}. It is however very well possible to implement this compatibility by taking the data scheme of these objects and serializing (or deserializing) them into JSON. With this feature available, \textbf{domain objects} could be easily transferred between the web application and web services that use the same schema. This way,  \textbf{native SIBs} following the HTTP-SIBs for the deserialization of the HTTP response would not longer be needed.

\paragraph{Using Domain Objects for APIs.}
\label{par:manualMapping}
The process described above would be useful in a scenario where both the web application and the web service use data schemes that are either equal or at least fairly compatible. The utility would most likely be limited to web services that a DIME user has implemented him/herself, specifically tailored to the web application at hand.\\
However, the DIME user could also adapt the DIME application to the web service that shall be accessed. With tools like \textbf{Postman}~\cite{Postman} or \textbf{Swagger Inspector}~\cite{SwaggerInspector} that are designed to quickly send HTTP requests towards an API to either test it or analyze its functionality, the DIME user could easily map the schema of the API within DIME. The process for this would begin with a test requests towards the desired API with one of the tools mentioned above. The response, in form of a JSON, would then be manually mapped to a \textbf{domain object} within DIME. This \textbf{domain object} could then be easily serialized and deserialized by the HTTP-DSL, resulting in a seamless connection between the web application and the web service.

\subsubsection{Automating Domain Object Mapping}
The \textbf{OpenAPI Specification} discussed in \autoref{subsubsec:AutoAPIAccess} allows for a very detailed documentation of RESTful APIs. One aspect of such a documentation is the description of the expected and returned JSON objects for each operation within the API. 

\begin{figure}[h]
	\begin{center}
		\includegraphics[scale=1]{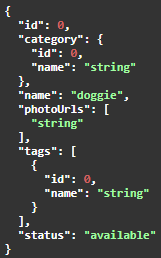}
		\caption{Description of Pet object from the SwaggerUI example.}
		\label{fig:PETSchema}
	\end{center}
\end{figure}

\paragraph{Pet Object Description.}
An example for such a description can be seen in \autoref{fig:PETSchema}. It shows a pet object that is used within the pet store API (from the example discussed in \autoref{par:petstore}) in form of JSON. A pet object contains basic \textit{id} and \textit{name} attributes, a \textit{photoUrls} attribute containing a list of Strings, a \textit{category} attribute containing a list of \classname{category} objects, a \textit{tags} attribute containing a list of \classname{tag} objects and finally a \textit{status} attribute holding an enum.

With a machine readable description of that kind, one could potentially automate the process that was manually done in \autoref{par:manualMapping}. With the help of the tools provided by Swagger, DIME could not only provide a library of HTTP SIBs (as described in \autoref{subsubsec:AutoAPIAccess}), but also provide \textbf{domain objects} that match the data schema of the desired web service.

\begin{figure}[h]
	\begin{center}
		\includegraphics[scale=1]{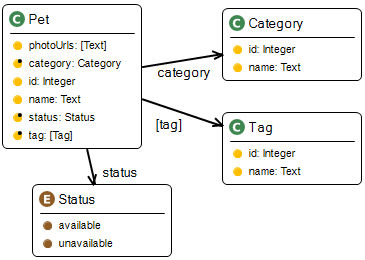}
		\caption{Domain object within DIME of a pet object.}
		\label{fig:PETDomainObject}
	\end{center}
\end{figure}

\paragraph{Transforming the Pet Object.}
An example for such a transformation from a JSON description to a DIME \textbf{domain object} for the pet object is depicted in \autoref{fig:PETDomainObject}. The object with all its attributes would be translated in conformity to the \textbf{Data Language} within DIME. The attributes that contain other objects would automatically trigger the creation of those objects as \textbf{domain objects} within DIME as well.

\newpage
\paragraph{Evaluating the Feature.}
This function could be used in two different scenarios:
\begin{itemize}
    \item The DIME user \textbf{uses the same data schema} as the web service that shall be accessed.
    \item The DIME user \textbf{has many data schemata} within the DIME application.
\end{itemize}
Both scenarios carry advantages and disadvantages. \\
In the first scenario, the clear disadvantage is that the data schema of the web service might not be optimal for the DIME application, thus limiting its functionality or cluttering it. The first scenario would however be quickly set up and have no compatibility issues with the web service that the DIME application is accessing.\\
The second scenario would have the advantage of a custom data schema for the application that would perfectly suit the needs of the DIME user. However, the disadvantage would be that \textbf{domain objects} of the web service would need to be translated (meaning their data transferred) to the \textbf{domain objects} used for the DIME application itself. This problem would also occur in the first scenario when accessing more than one web service, as multiple web services almost never share the same data schemata.\\
Thus, the need to transfer data between \textbf{domain objects} is prevalent almost all the time. Luckily, transferring data from one \textbf{domain object} to another within DIME is not a difficult feat. The DIME user could quickly model processes that map \textbf{domain objects} from and to each other, meaning that the overall solution of providing the data schemata of web services is very useful. When comparing the task of mapping \textbf{domain objects} to the one of writing custom JSON serializers and deserializers for \textbf{domain objects}, the former is more efficient and provides fewer chances for bugs and errors, as no code is written by hand. \\[5mm]

Overall, all these further extensions to the HTTP-DSL illustrate that it can act as a foundation for many further ideas. This fits perfectly into the mindset of the LDE paradigm, discussed in \autoref{subsubsec:LDE}. DIME is upgraded and further extended with the integration of the HTTP-DSL, which is yet another DSL that is added to the pool of DSLs that make DIME what it is today. However, it does not stop there, as it can be argued that the HTTP-DSL can be used to generate further DSLs down the road, such as the REST layer. In the original paper introducing LDE \cite{LDE} it was written that
\begin{quote}
    "[T]he interplay between the involved DSLs is realized in a service-oriented fashion. This eases a product line approach and system evolution by allowing to introduce and exchange entire DSLs within corresponding Mindset-Supporting Integrated Development Environments (mIDEs)."\cite{LDE}
\end{quote}
The HTTP-DSL project is exactly that. DIME, with the help of the Generic SIBs plugin, allowed for a fairly simple introduction of a completely new DSL that can now be used to address a need within DIME that was not addressed before, extending the mIDE by another service. The HTTP-DSL integration can be argued to be a proof that this concept of modular extensions to a system by DSLs is very well possible within DIME and it became easier than it has ever been before with the introduction of the Generic SIBs plugin.

\attachmentmatter

\printbibliography

@book{Fowler10,
  author = {Fowler, Martin},
  isbn = {978-0-321-71294-3},
  publisher = {Addison-Wesley},
  title = {Domain-Specific Languages},
  username = {flint63},
  year = 2010
}

@ONLINE{HTTP1,
  title = {W3 HTTP Statistics},
  author = {W3techs},
  url = {https://w3techs.com/technologies/comparison/ce-http2,ce-http3},
  urldate = {2021-12-11}
}

@ONLINE{HTTPCLIENT,
  title = {Http Client Repository},
  author = {Apache},
  url = {https://mvnrepository.com/artifact/org.apache.httpcomponents/httpclient/4.5.13},
  urldate = {2021-12-11}
}

@ONLINE{RFC1738,
  title = {IETF RFC 1738},
  author = {IETF},
  url ={https://datatracker.ietf.org/doc/html/rfc1738},
  urldate = {2021-12-11}
}

@ONLINE{TOPDOM,
  title = {IETF Top Level Domain Name Specification},
  author = {IETF},
  url = {https://tools.ietf.org/id/draft-liman-tld-names-00.html},
  urldate = {2021-12-11}
}

@ONLINE{RFC5854,
  title = {IETF IPv6 Specification},
  author = {IETF},
  url = {https://datatracker.ietf.org/doc/html/rfc5954},
  urldate = {2021-12-11}
}

@ONLINE{RFC1883,
  title = {First introduction of IPv6},
  author = {IETF},
  url = {https://datatracker.ietf.org/doc/html/rfc1883},
  urldate = {2021-01-02}
}

@ONLINE{projectDescription,
  title = {Cinco Product Definition},
  author = {CS Chair 5, TU-Dortmund},
  url = {https://gitlab.com/scce/cinco/-/wikis/Cinco-Product-Definition},
  urldate = {2021-12-17}
}

@ONLINE{accuwheather,
  title = {Accuweather Weather API},
  url = {https://developer.accuweather.com/},
  urldate = {2021-12-19}
}

@ONLINE{accuwheatherLoc,
  title = {Accuweather Locations API},
  author = {Accuweather},
  url = {https://developer.accuweather.com/accuweather-locations-api/apis},
  urldate = {2021-12-19}
}

@ONLINE{accuwheatherCC,
  title = {Accuweather Current Conditions API},
  author = {Accuweather},
  url = {https://developer.accuweather.com/accuweather-current-conditions-api/apis},
  urldate = {2021-12-19}
}

@ONLINE{dimeInstall,
  title = {DIME Wiki instructions and download page},
  author = {CS Chair 5, TU-Dortmund},
  url = {https://scce.gitlab.io/dime/content/introduction/},
  urldate = {2021-12-19} 
}

@ONLINE{webService,
  title = {W3C description of Web Services Architecture},
  author = {W3C},
  url = {https://www.w3.org/TR/ws-arch/},
  urldate = {2021-12-28}
}

@book{PoppendieckPoppendieck03,
  address = {Boston, MA},
  author = {Poppendieck, Mary and Poppendieck, Tom},
  isbn = {978-0-321-15078-3},
  publisher = {Addison-Wesley},
  title = {Lean Software Development: An Agile Toolkit},
  url = {https://www.safaribooksonline.com/library/view/ean-software-development/0321150783/},
  username = {flint63},
  year = 2003
}

@Inbook{LDE,
author="Steffen, Bernhard
and Gossen, Frederik
and Naujokat, Stefan
and Margaria, Tiziana",
title="Language-Driven Engineering: From General-Purpose to Purpose-Specific Languages",
bookTitle="Computing and Software Science: State of the Art and Perspectives",
year="2019",
publisher="Springer International Publishing",
address="Cham",
pages="311--344",
doi="10.1007/978-3-319-91908-9_17"
}

@mastersthesis{Kopetz2014,
	author = {Dawid Kopetzki},
	month = jun,
	school = {TU Dortmund},
	title = {{M}odel-based generation of graphical editors on the basis of abstract meta-model specifications},
	type = {Master thesis},
	year = {2014}}

@mastersthesis{Thoene,
	author = {Jonathan Th{\"o}ne},
	school = {Technische Universit{\"a}t Dortmund},
	title = {Entwicklung einer Plugin-Architektur zur Erweiterung von DIME-Prozessen um externe SIB-Typen},
	type = {Master Thesis (in german)},
	year = {2020}}

@ONLINE{Xtext,
  title = {Official Xtext Webpage},
  author = {Oracle},
  url = {https://www.eclipse.org/Xtext/},
  urldate = {2021-12-30}
}

@ONLINE{EMF,
  title = {Official EMF Webpage},
  author = {Oracle},
  url = {https://www.eclipse.org/modeling/emf/},
  urldate = {2021-12-30}
}

@ONLINE{intellij,
  title = {Official webpage of IntelliJ IDEA},
  author = {Jetbrains},
  url ={https://www.jetbrains.com/de-de/idea/},
  urldate = {2021-12-31}
}

@ONLINE{vscode,
  title = {Official web page of Visual Studio Code by Microsoft.},
  author = {Microsoft},
  url = {https://code.visualstudio.com/},
  urldate = {2021-12-31}
}

@ONLINE{uml,
  title = {UML specifications by the OMG.} ,
  author = {Object Management Group},
  url ={https://www.omg.org/spec/UML/2.5.1/About-UML/},
  urldate = {2022-01-01}
}

@ONLINE{XP,
  title = {A lecture about eXtreme Programming.},
  author = {Terrence Parr},
  location = {University of San Francisco},
  url ={https://www.cs.usfca.edu/~parrt/course/601/lectures/xp.html},
  urldate = {2022-01-01}
}

@ONLINE{CICD,
  title = {A blog post explaining CI/CD.},
  author = {Isaac Sacolick},
  url = {https://www.infoworld.com/article/3271126/what-is-cicd-continuous-integration-and-continuous-delivery-explained.html},
  urldate = {2022-01-01}
}

@Inbook{Margaria2020,
author="Margaria, Tiziana
and Steffen, Bernhard",
editor="Tatnall, Arthur",
title="eXtreme Model-Driven Development Technologies as a Hands-On Approach to Software Development Without Coding",
bookTitle="Encyclopedia of Education and Information Technologies",
year="2020",
publisher="Springer International Publishing",
address="Cham",
pages="1--19",
doi="10.1007/978-3-319-60013-0_208-1"
}

@ONLINE{ietf,
  title = {Official IETF Website.},
  author = {IETF},
  url = {https://www.ietf.org/},
  urldate = {2022-01-02}
}

@ONLINE{w3,
  title = {Official World Wide Web Consortium Website.},
  author = {W3C},
  url = {https://www.w3.org/},
  urldate = {2022-01-02}
}

@ONLINE{rfc2818,
  title = {RFC 2818 specifications.},
  author = {IETF},
  url = {https://datatracker.ietf.org/doc/html/rfc2818},
  urldate = {2022-01-02}
}

@ONLINE{angular,
  title = {Angular Homepage.},
  author = {Metaverse},
  url = {https://angular.io/},
  urldate = {2022-01-02}
}

@ONLINE{OAuth2,
  title = {OAuth 2.0 Homepage.},
  author = {Aaron Parecky},
  url = {https://oauth.net/2/},
  urldate = {2022-01-06}
}

@ONLINE{basicAuth,
  title = {RFC 1945.},
  author = {IETF},
  url = {https://datatracker.ietf.org/doc/html/rfc1945},
  urldate = {2022-01-06}
}

@book{userstories1,
author = {Cohn, Mike},
title = {User Stories Applied: For Agile Software Development},
year = {2004},
isbn = {0321205685},
publisher = {Addison Wesley Longman Publishing Co., Inc.},
address = {USA}
}

@ONLINE{userstories2,
  title = {Project Advantages of User Stories as Requirements},
  author = {Cohn, Mike},
  url = {https://www.mountaingoatsoftware.com/articles/advantages-of-user-stories-for-requirements},
  urldate = {2022-01-08}
}

@ONLINE{ITUGlossary,
  title = {ITU-T Rec. Q.1290 (05/98) Glossary of terms used in the definition of
intelligent networks},
  author = {International Telecommunication Union},
  url = {https://www.itu.int/ITU-T/recommendations/rec.aspx?rec=3233&lang=en},
  urldate = {2022-01-08}
}

@ONLINE{swagger,
  title = {Official Swagger Project Website.},
  author = {Swagger},
  url = {https://swagger.io/},
  urldate = {2022-01-12}
}

@ONLINE{OpenAPI2,
  title = {Difference between Swagger and OpenAPI.},
  author = {Swagger},
  url = {https://swagger.io/blog/api-strategy/difference-between-swagger-and-openapi/},
  urldate = {2022-01-12}
}

@ONLINE{smartbear,
  title = {Official Website of SmartBear.},
  author = {SmartBear},
  url ={https://smartbear.com/},
  urldate = {2022-01-12}
}

@ONLINE{Swaggerui,
  title = {SwaggerUI Pet store example.},
  author = {Swagger},
  url = {https://petstore.swagger.io/},
  urldate = {2022-01-12}
}

@ONLINE{SwaggerInspector,
  title = {SwaggerInspector tool.},
  author = {Swagger},
  url = {https://inspector.swagger.io/builder?utm_campaign=openapi_swagger_difference&utm_medium=blog&utm_source=swagger},
  urldate = {2022-01-12}
}

@ONLINE{Postman,
  title = {Official Postman Website.},
  author = {Postman},
  url = {https://www.postman.com/},
  urldate = {2022-01-12}
}

@ONLINE{jetbrains,
  title = {Official Jetbrains Website.},
  author = {Jetbrains},
  url = {https://www.jetbrains.com/},
  urldate = {2022-01-15}
}

@INPROCEEDINGS{9568378,  author={Jobish John and  Amrita Ghosal and Tiziana Margaria and Dirk Pesch},  booktitle={2021 Forum on specification   Design Languages (FDL)},   
title = {{D}SLs for {M}odel {D}riven {D}evelopment of {S}ecure {I}nteroperable {A}utomation {S}ystems with {E}dgeX {F}oundry},   year={2021},  
volume={},  
number={},  
pages={1-8},  
doi={10.1109/FDL53530.2021.9568378}}

@InProceedings{industrial3,
author="John, Jobish
and Ghosal, Amrita
and Margaria, Tiziana
and Pesch, Dirk",
%editor="Margaria, Tiziana
%and Steffen, Bernhard",
title="DSLs and Middleware Platforms in a Model-Driven Development Approach for Secure Predictive Maintenance Systems in Smart Factories",
booktitle="Proc. ISoLA 2021, Leveraging Applications of Formal Methods, Verification and Validation, LNCS vol. 13036",
year="2021",
publisher="Springer International Publishing",
pages="146--161",
abstract="In many industries, traditional automation systems (operating technology) such as PLCs are being replaced with modern, networked ICT-based systems as part of a drive towards the Industrial Internet of Things (IIoT). The intention behind this is to use more cost-effective, open platforms that also integrate better with an organisation's information technology (IT) systems. In order to deal with heterogeneity in these systems, middleware platforms such as EdgeX Foundry, IoTivity, FI-WARE for Internet of Things (IoT) systems are under development that provide integration and try to overcome interoperability issues between devices of different standards. In this paper, we consider the EdgeX Foundry IIoT middleware platform as a transformation engine between field devices and enterprise applications. We also consider security as a critical element in this and discuss how to prevent or mitigate the possibility of several security risks. Here we address secure data access control by introducing a declarative policy layer implementable using Ciphertext-Policy Attribute-Based Encryption (CP-ABE). Finally, we tackle the interoperability challenge at the application layer by connecting EdgeX with DIME, a model-driven/low-code application development platform that provides methods and techniques for systematic integration based on layered Domain-Specific Languages (DSL). Here, EdgeX services are accessed through a Native DSL, and the application logic is designed in the DIME Language DSL, lifting middleware development/configuration to a DSL abstraction level. Through the use of DSLs, this approach covers the integration space domain by domain, technology by technology, and is thus highly generalizable and reusable. We validate our approach with an example IIoT use case in smart manufacturing."
%isbn="978-3-030-89159-6"
}

@ONLINE{maven,
  title = {Official maven Website.},
  author = {Maven},
  url = {https://maven.apache.org/},
  urldate = {2022-01-17}
}

@ONLINE{RCP,
  title = {Official RCP Website.},
  author = {Eclipse},
  url = {https://wiki.eclipse.org/Rich_Client_Platform},
  urldate = {2022-01-17}
}

@preamble{"\def\Nst#1{$#1^{st}$}" #
   "\def\Nnd#1{$#1^{nd}$}" #
   "\def\Nrd#1{$#1^{rd}$}" #
   "\def\Nth#1{$#1^{th}$}"}

@string{cs = {Computing Surveys}}

@string{it = {Dept. of Information Technology, Uppsala University, Sweden}}

@string{lncs = {Lecture Notes in Computer Science}}

@inproceedings{DBLP:conf/ifm/MargariaS19,
	author = {Tiziana Margaria and Alexander Schieweck},
	bibsource = {dblp computer science bibliography, https://dblp.org},
	booktitle = { Proc.  15th Int.  Conf. on Integrated Formal Methods (IFM 2019)},
	pages = {3--24},
	publisher = {Springer},
	series = {LNCS},
	title = {The Digital Thread in Industry 4.0},
	volume = {11918},
	location = {Bergen, Norway},
	year = {2019}
	}

@book{BecAnd2004,
	author = {Beck, Kent and Andres, Cynthia},
	publisher = {Addison-Wesley Professional},
	title = {{E}xtreme programming explained: embrace change},
	year = {2004}}

@inproceedings{BFKLNN2016,
	author = {Steve Bo{\ss}elmann and Markus Frohme and Dawid Kopetzki and Michael Lybecait and Stefan Naujokat and Johannes Neubauer and Dominic Wirkner and Philip Zweihoff and Bernhard Steffen},
	booktitle = {Proc. of the 7th Int. Symp. on Leveraging Applications of Formal Methods, Verification and Validation, Part II (ISoLA 2016)},
	doi = {10.1007/978-3-319-47169-3\_60},
	pages = {809--832},
	publisher = {Springer},
	series = {LNCS},
	title = {{DIME}: {A} {P}rogramming-{L}ess {M}odeling {E}nvironment for {W}eb {A}pplications},
	volume = {9953},
	year = {2016}}

@book{BrCaWi2012,
	author = {Marco Brambilla and Jordi Cabot and Manuel Wimmer},
	doi = {10.2200/S00441ED1V01Y20120-8SWE001},
	publisher = {Morgan \& Claypool},
	title = {{M}odel-{D}riven {S}oftware {E}ngineering in {P}ractice},
	year = {2012}}

@phdthesis{Fielding2000,
	author = {Roy Thomas Fielding},
	school = {University of California, Irvine},
	title = {{A}rchitectural {S}tyles and the {D}esign of {N}etwork-based {S}oftware {A}rchitectures},
	url = {http://www.ics.uci.edu/~fielding/pubs/dissertation/top.htm},
	year = {2000}}

@article{NaLyKS2017,
	author = {Stefan Naujokat and Michael Lybecait and Dawid Kopetzki and Bernhard Steffen},
	doi = {10.1007/s10009-017-0453-6},
	journal = {Software Tools for Technology Transfer},
	pages = {327-354},
	title = {{CINCO}: {A} {S}implicity-{D}riven {A}pproach to {F}ull {G}eneration of {D}omain-{S}pecific {G}raphical {M}odeling {T}ools},
	year = {2017}}

\newpage
\begin{appendices}

\fontsize{8}{12}\selectfont

\section{DSL Example}
\label{app:dslExample}
\begin{lstlisting}[
style=HTTP,
caption={DSL Example the full spectrum of the DSL.}
]
http{
	name "SomeApiRequest"
	url server http://www.someurl.com path somepath/somemore
	type GET
	param "staticParamKey" : "staticParamValue"
	param input $VariableParamKey : input $VariableParamValue
	header Authorization : "Base cGFzc3dvcmQ="
	body{
		contentType application/json
		payload "somePayload"
		entityType StringEntity
	}
	customize{
		ProxyServer name "someserver.to" port 80 
		basicAuth username : "user" password : "password"
		timeout 9000
	}

}
\end{lstlisting}

\section{HTTP DSL Code}
\label{appendix:dslCode}
\begin{lstlisting}[
label={app:DSL},
style=xtext,
caption={Full Xtext code for designing the HTTP DSL.}
]
grammar info.scce.cinco.http.Http with org.eclipse.xtext.common.Terminals

import "http://www.eclipse.org/emf/2002/Ecore" as ecore 

generate http "http://www.scce.info/cinco/http/Http"

HttpFile: {HttpFile}
	messages += HttpMessage+
;

HttpMessage: 
	'http' '{'
		'name' name = STRING
		'url' url = AbstractUrl
		'type' requestMethod = RequestMethod
		query += Param*
		headers += Header*
		body = Body?
		returnValue = ReturnValue?
		customization = Customization?
		'}'
;

//Basic terminals
terminal fragment DIGIT:('0'..'9');
@Override
terminal INT returns ecore::EInt: DIGIT+ ; 
terminal fragment UCHAR: (ALPHADIGIT|"$" | "-" | "_" | "." | "+"| "!" | "*" | "'" | "(" | ")" | ",");
terminal fragment ALPHADIGIT: ALPHA | DIGIT;
terminal fragment ALPHAUP:('A'..'Z');
terminal fragment ALPHA: ('a'..'z')|ALPHAUP;


//Variables
AbstractVariable: (Inputvariable|Environmentvariable);
Inputvariable: 'input' value = INPUTVARIABLENAME;
Environmentvariable: 'environment' value = ENVIRONMENTVARIABLENAME;
terminal ENVIRONMENTVARIABLENAME: (ALPHAUP|'_')+;
terminal INPUTVARIABLENAME: '$' (ALPHADIGIT|('-')|('.')|('_')|('~')|('/'))+ ;


AbstractUrl: 'server' (server = SpecifiedServer  | server = AbstractVariable)
				('path' (path = SpecifiedPath | path = AbstractVariable))?
;
SpecifiedServer: value = SERVER;
SpecifiedPath: value=PATH;

//SERVER
terminal SERVER:(PROTOCOL)? (CREDENTIAL ':' CREDENTIAL '@')? HOST ;
terminal fragment PROTOCOL: 'http''s'? ':' '//';
terminal fragment HOST: (HOSTNAME | HOSTIPV4 | HOSTIPV6) (':' INT)?;
terminal fragment HOSTNAME: (DOMAINLABEL '.')* TOPLABEL;
terminal fragment DOMAINLABEL: ALPHADIGIT | (ALPHADIGIT (ALPHADIGIT | "-")* ALPHADIGIT);
terminal fragment TOPLABEL: ALPHA (ALPHADIGIT | '-')* ALPHADIGIT;
terminal fragment HOSTIPV4: INT '.' INT '.' INT '.' INT;
terminal fragment HOSTIPV6:'[' HEXPART (':'HOSTIPV4)?']';
terminal fragment HEXPART: HEXSEQ | (HEXSEQ '::' (HEXSEQ)? )|('::'(HEXSEQ)?);
terminal fragment HEXSEQ: HEX+ (':' HEX+)*;
terminal fragment HEX: ('0'..'9')|('a'..'f')|('A'..'F');
terminal fragment CREDENTIAL:(UCHAR | ";" | "?" | "&" | "=")+;
//PATH
terminal PATH: SEGMENT ('/' SEGMENT)*; 
terminal fragment SEGMENT:(UCHAR |";" | ":" | "@" | "&" | "=" )+; 

//Request type
RequestMethod: value=TYPESTRING;
enum TYPESTRING: GET | POST | PUT | DELETE;

//Param
Param: 'param' (key=ParamKey | key=AbstractVariable) ':' (value=ParamValue|value=AbstractVariable);
ParamKey: value=STRING;
ParamValue: value=STRING;

//Header

Header: 'header' 
	(key = AbstractHeaderKey | key=AbstractVariable) ':' 
	(value = HeaderValue | value=AbstractVariable);
AbstractHeaderKey:(HeaderKey | CustomHeaderKey);
HeaderKey: value=HEADERKEYSTRING;
CustomHeaderKey: value=STRING;
HeaderValue: value=STRING;
enum HEADERKEYSTRING: 	
	
						AGE="Age"|ALLOW='Allow'|ALTERNATEPROTOCOL="Alternate-Protocol"|CLIENTDATE="Client-Date"|CLIENTPEER="Client-Peer"|
						CONNECTION="Connection"|AUTHORIZATION="Authorization"|RETRYAFTER="Retry-After" | ACCESSCONTROLALLOWCREDENTIALS="Access-Control-Allow-Credentials" 
						| ACCESSCONTROLALLOWHEADERS="Access-Control-Allow-Headers"| ACCESSCONTROLALLOWMETHODS="Access-Control-Allow-Methods" | ACCESSCONTROLALLOWORIGIN="Access-Control-Allow-Origin"
						| ACCESSCONTROLEXPOSEHEADERS="Access-Control-Expose-Headers" | ACCESSCONTROLMAXAGE="Access-Control-Max-Age" | ACCEPTRANGED="Accept-Ranges" | 
						CACHECONTROL="Cache-Control"| CLIENTRESPONSENUM="Client-Response-Num" | CONTENTDISPOSITION ="Content-Disposition"| CONTENTENCODING="Content-Encoding"|
						CONTENTLANGUAGE="Content-Language"|CONTENTLENGTH="Content-Length"|CONTENTLOCATION="Content-Location"|CONTENTMD5="Content-MD5"|CONTENTRANGE="Content-Range"|
						CONTENTTYPE="Content-Type"|DATE="Date"|ETAG="ETag"|EXPIRES="Expires"|HTTP="HTTP"|KEEPALIVE="Keep-Alive"| LASTMODIFIED="Last-Modified"|LINK="Link"|LOCATION="Location"|
						P3P="P3P"|PRAGMA="Pragma"|PROXYAUTHENTICATE="Proxy-Authenticate"|PROXYCONNECTION="Proxy-Connection"|REFRESH="Refresh"|SERVER="Server"| SETCOOKIE="Set-Cookie"
						|STATUS="Status"|STRICTTRANSPORTSECURITY="Strict-Transport-Security"|TIMINGALLOWORIGIN="Timing-Allow-Origin"|TRAILER="Trailer"|TRANSFERENCODING="Transfer-Encoding"|
						UPGRADE="Upgrade"|VARY="Vary"|VIA="Via"|WARNING="Warning"|WWWAUTHENTICATE="WWW-Authenticate"|XASPNETVERSION="X-Aspnet-Version="
;

//Body
Body: 'body' '{'
	'contentType' (contentType = ContentType | contentType = AbstractVariable | contentType = CustomContentType)
	'payload' (payload = BodyPayload | payload = AbstractVariable)
	'entityType' entityType = ENTITYTYPE
	'}'
;
BodyPayload: value=STRING ;

ContentType: value = CONTENTTYPETERM;
CustomContentType: value = STRING;

enum ENTITYTYPE: STRING="StringEntity"|FILE="FileEntity"|BYTEARRAY="ByteArrayEntity"|STREAM="InputStreamEntity";
enum CONTENTTYPETERM: TEXTPLAIN="text/plain"|TEXTCSS="text/css"|TEXTHTML="text/html"|TEXTJ="text/javascript"|
				IMAGEAPNG="image/apng"|IMAGEAVIF="image/avif"|IMAGEGIF="image/gif"|IMAGEJPEG="image/jpeg"|IMAGEPNG="image/png"|IMAGESVG="image/svg+xml"|IMAGEWEBP="image/webp"|
				AUDIOWAV="audio/wav"|AUDIOWEBM="audio/webm"|VIDEOWEBM="video/webm"|VIDEOMP4="video/mp4"|
				APPLICATIONJSON="application/json"|APPLICATIONPDF="application/pdf"|APPLICATIONZIP="application/zip"|APPLICATIONRAR="application/vnd.rar"|APPLICATIONTAR="application/x-tar"|
				APPLICATIONXML="application/xml"
;

ReturnValue: 'response' 
	'type' (contentType = ContentType | contentType = AbstractVariable | contentType = CustomContentType)
	'return' returnForm=RETURNFORM
;
enum RETURNFORM: ENTIRE_MESSAGE="Entire message"|PAYLOAD_ONLY="Payload only";

Customization:{Customization} 
				'customize' '{'
				proxyServer = ProxyServer?
				basicAuth = BasicAuth?
				timeout= Timeout?
	'}'
;
//Proxy
ProxyServer: 'ProxyServer' 'name' (proxyName = ProxyEndpoint | proxyName = AbstractVariable) 'port' port = INT;
ProxyEndpoint: value = STRING;
	
//Auth				
BasicAuth: 'basicAuth' 
			'username'':' username = (AuthUsername|AbstractVariable)
			'password'':' password = (AuthPassword|AbstractVariable)
			;
AuthUsername:value=STRING;
AuthPassword:value=STRING;

//Timeout
Timeout: 'timeout' value= INT;
\end{lstlisting}

\section{CustomResponseHandler}
\label{app:CRH}
\begin{lstlisting}[
style=java,
caption={Full CustomResponseHandler Java Code.}
]
package testApp;

import org.apache.http.Header;
import org.apache.http.HttpEntity;
import org.apache.http.HttpResponse;
import org.apache.http.StatusLine;
import org.apache.http.client.ClientProtocolException;
import org.apache.http.client.ResponseHandler;
import org.apache.http.util.EntityUtils;
import testApp.ResponseObject;

import java.io.IOException;

public class CustomResponseHandler implements ResponseHandler<ResponseObject> {
   @Override
    public ResponseObject handleResponse(HttpResponse response){

        StatusLine statusLine = response.getStatusLine();
        HttpEntity entity = response.getEntity();
        String payload = EntityUtils.toString(entity, "UTF-8");

        int statusCode = statusLine.getStatusCode();
        Header locationHeader = response.getFirstHeader("Location");
        switch (statusCode){
            case 300:   //Multiple Choices
                        //try again with same Method and new uri.
                if(locationHeader != null) {
                    return new ResponseObject(payload, statusCode, false, true, 
                    locationHeader.getValue(),RequestType.AS_BEFORE);
                }else{
                    return new ResponseObject(payload, statusCode, false, false, 
                    null,RequestType.AS_BEFORE);
                }
            case 305:   //Redirect, Proxy, Temporary Redirect, Permanent Redirect
            case 301:   //try again with same Method and new uri.
            case 307:
            case 308:
                    return new ResponseObject(payload, statusCode, false, true,
                    locationHeader.getValue(),RequestType.AS_BEFORE);
            case 303:   //See Other
            case 302:   // try again with new Method and new uri.
                    return new ResponseObject(payload, statusCode, false,true,
                    locationHeader.getValue(),RequestType.GET);
            case 400:
            case 401:       
            case 403:
            case 404:
            case 405:
            case 407:
            case 409:
            case 410:
            case 411:
            case 412:
            case 414:
            case 415:
            case 416:
            case 417:
            case 421:
            case 422:
            case 424:
            case 426:
            case 428:
            case 431:
            case 451:
            case 418:
            case 420:
            case 444:
            case 449:
            case 499:
            case 500:
            case 501:
            case 502:
            case 503:
            case 504:
            case 505:
            case 506:
            case 508:
            case 510:
            case 511:
                return new ResponseObject(payload, statusCode, false,false, 
                null, RequestType.AS_BEFORE );
            case 408:   //Request Timeout
                        //Erneut versuchen
                return new ResponseObject(payload, statusCode, false,true, 
                null,RequestType.AS_BEFORE);
            case 413:   //Payload Too Large
                        //retry after time specified in "Retry-After" header or after 5 seconds
                Header timeoutHeader = response.getFirstHeader("Retry-After");
                if(timeoutHeader != null){
                    try {
                        wait(new Integer(timeoutHeader.getValue()));
                        return new ResponseObject(payload, statusCode, false,true,
                        null,RequestType.AS_BEFORE);
                    } catch (InterruptedException e) {
                        e.printStackTrace();
                    }
                }else{
                    try {
                        wait(5000);
                        return new ResponseObject(payload, statusCode, false,true,
                        null,RequestType.AS_BEFORE);
                    } catch (InterruptedException e) {
                        e.printStackTrace();
                    }
                }
                break;
            case 423:   
            case 425:   
            case 429:
            case 507:
            case 509:
                try {
                    wait(5000);
                    return new ResponseObject(payload, statusCode, false,true,
                    null,RequestType.AS_BEFORE);
                } catch (InterruptedException e) {
                    e.printStackTrace();
                }
                break;
            default:    //20x
                        //everything worked as intended
                return new ResponseObject(payload, statusCode, true, false, 
                null, RequestType.AS_BEFORE);

        }

        throw new IllegalStateException("Unexpected errorcode occurred.");
    }
}
\end{lstlisting}

\section{REST Layer DSL description file}
\label{appendix:RestDSLFile}
\begin{lstlisting}[
style=HTTP,
caption={Full HTTP-DSL code for REST.}
]
http{
	name "GET Request"
	url server input $url path input $path
	type GET
	param  input $param1Key : input $param1Value
	param  input $param2Key : input $param2Value
	header input $header1Name : input $header1Value
	header input $header2Name : input $header2Value
}

http{
	name "POST Request"
	url server input $url path input $path
	type POST
	param  input $param1Key : input $param1Value
	param  input $param2Key : input $param2Value
	header input $header1Name : input $header1Value
	header input $header2Name : input $header2Value
	body{contentType input $BodyContentType payload input $BodyPayload entityType StringEntity}
	}
	
http{
	name "PUT Request"
	url server input $url path input $path
	type PUT
	param  input $param1Key : input $param1Value
	param  input $param2Key : input $param2Value
	header input $header1Name : input $header1Value
	header input $header2Name : input $header2Value
	body{contentType input $BodyContentType payload input $BodyPayload entityType StringEntity}
}

http{
	name "DELETE Request"
	url server input $url path input $path
	type DELETE
	param  input $param1Key : input $param1Value
	param  input $param2Key : input $param2Value
	header input $header1Name : input $header1Value
	header input $header2Name : input $header2Value
}
\end{lstlisting}

\end{appendices}


\end{document}